\documentclass[11pt,a4paper]{article}
\pdfoutput=1
\usepackage{jcappub}
\usepackage[utf8]{inputenc}
\usepackage{lmodern}
\usepackage[T1]{fontenc}

\usepackage{amsfonts}
\usepackage{mathtools}
\DeclareMathAlphabet{\mathsfi}{OT1}{cmss}{m}{sl}
\DeclareMathAlphabet{\mathbfi}{OML}{cmm}{b}{it}
\usepackage{bbm}
\usepackage{mathrsfs}
\usepackage{upgreek}

\newcommand{\total}{\mathop{}\!\mathrm{d}}
\newcommand{\eqend}[1]{\,\mathrm{#1}}

\renewcommand{\vec}[1]{\mathbfi{#1}}

\newcommand{\laplace}{\mathop{}\!\bigtriangleup}

\newcommand{\bigo}[1]{\mathcal{O}\left({#1}\right)}

\newcommand{\lie}{\mathscr{L}}

\newcommand{\bnabla}{\boldsymbol{\nabla}}

\newenvironment{equations}[1][]{\subequations\ifx\relax#1\relax\else\label{#1}\fi\align\ignorespaces}{\endalign\ignorespacesafterend\endsubequations}
\makeatletter
\def\@spliteq#1{\begin{equation}\begin{split}#1\end{split}\end{equation}}
\def\splitequation{\collect@body\@spliteq}

\makeatother

\let\originalleft\left
\let\originalright\right
\renewcommand{\left}{\mathopen{}\mathclose\bgroup\originalleft}
\renewcommand{\right}{\aftergroup\egroup\originalright}

\usepackage{xspace}

\newcommand{\ie}{\textit{i.\,e.}\xspace}
\newcommand{\eg}{\textit{e.\,g.}\xspace}

\begin{document}

\title{Cosmological Perturbations and Invariant Observables in Geodesic Lightcone Coordinates}

\author[a]{Markus B. Fr{\"o}b}
\author[b]{and William C. C. Lima}

\affiliation[a]{Institut f\"ur Theoretische Physik, Universit{\"a}t Leipzig,\\ Br{\"u}derstra{\ss}e 16, 04103 Leipzig, Germany}
\affiliation[b]{Department of Mathematics, University of York,\\ Heslington, York, YO10 5DD, United Kingdom}

\emailAdd{mfroeb@itp.uni-leipzig.de}
\emailAdd{william.correadelima@york.ac.uk}

\abstract{We consider a recent approach to the construction of gauge-invariant relational observables in gravity in the context of cosmological perturbation theory. These observables are constructed using a field-dependent coordinate system, which we take to be geodesic lightcone coordinates. We show that the observables are gauge-independent in the fully non-linear theory, and that they have the expected form when one adopts the geodesic lightcone gauge for the metric. We give explicit expressions for the Sasaki--Mukhanov variable at linear order, and the Hubble rate --- as measured both by geodesic observers and by observers co-moving with the inflaton --- to second order. Moreover, we show that the well-known linearised equations of motion for the Sasaki-Mukhanov variable and the scalar constraint variables follow from the gauge-invariant Einstein's equations.}



\maketitle

\section{Introduction}                                                                                                                              %
\label{sec:introduction}                                                                                                                            %

The increasing precision in current~\cite{conley_etal_ajs_2011,planck_2018a,planck_2018b,planck_2018c} and planned experiments~\cite{lsstdark_energy_science_arxiv_2012,euclid_th_working_grp_lrr_2013,desi_arxiv_2016,cmb-s4_arxiv_2016} to map the universe at its largest scales has renewed the interest in non-linear effects in relativistic cosmology. The standard way to approach these effects is to treat inhomogeneities as perturbations over a Friedmann-Lema\^{\i}tre-Robertson-Walker (FLRW) spacetime, and exploit the background symmetries to decompose the perturbations in the metric and possibly other fields. At linear order, the theory has correctly explained the anisotropies of the cosmic microwave background and the formation of structure at large scales, see for example Ref.~\cite{dodelson_cosmology_book}.

To extend the theory of cosmological perturbations to higher orders, one faces the challenge of identifying and solving for the relevant gauge-invariant observables of the theory. When treating gravity perturbatively, the invariance of the underlying gravitational theory under diffeomorphisms translates into a gauge symmetry for the metric perturbation. The fact that this symmetry acts on the spacetime manifold by displacing its points implies that local fields, \ie, fields defined at a fixed point of the background spacetime, are gauge dependent. Consequently, local fields cannot describe true observables of the theory. Indeed, even though it is always possible to find a complete set of local observables in linearised gravity~\cite{froeb_hack_higuchi_jcap_2017,froeb_hack_khavkine_cqg_2018,khavkine_cqg_2019}, it has been shown that gauge-invariant observables in the full theory are necessarily non-local~\cite{torre_prd_1993,giddings_marolf_hartle_prd_2006,khavkine_cqg_2015}.

A possible way to construct invariant observables in General Relativity is to define them in a relational way. In this approach, observables are given by fields evaluated at spacetime points that are not fixed, but instead determined by the value that other dynamical fields of the theory assume there. Hence, in the relational approach one defines both the quantity that is measured and the frame with respect to which the measurement is performed. This approach goes back a long way in the General Relativity literature~\cite{komar_pr_1958,bergmann_komar_prl_1960,bergmann_rmp_1961}, see Ref.~\cite{tambornino_sigma_2012} for a recent review. In general, the relational approach relies on taking scalar fields constructed from the (gauge-dependent) degrees of freedom of the system, and employing them as coordinates. These fields can be constructed as geometrical scalars (such as curvature scalars), in which case one needs a background spacetime that is inhomogeneous enough such that its points may be discriminated by the value of these scalars. If this does not hold, another possibility is to add scalar fields by hand, such as the famous Brown-Kucha\v{r} dust model~\cite{brown_kuchar_prd_1995}. These additional dust fields, however, change the physics of the system, \ie, modify the dynamics of gauge-invariant observables such as the Bardeen potentials~\cite{bardeen_prd_1980} and the Sasaki--Mukhanov variable~\cite{sasaki_ptp_1986, mukhanov_zetf_1988}. Indeed, it has been shown that the equations of motion for these observables contain extra terms (with respect to their standard forms) due to the presence of the dust reference frame~\cite{giesel_et_al_cqg_2010a,giesel_et_al_cqg_2010b,giesel_et_al_arxiv_2020}.

An alternative way to obtain these field-dependent coordinates is to construct them from certain components of the spacetime metric (or other dynamical fields that are present). This avoids the introduction of new degrees of freedom to the system, but results in coordinates that depend non-locally on the metric. A concrete way to build the field-dependent coordinates was proposed in Ref.~\cite{brunetti_etal_jhep_2016} and further developed in Refs.~\cite{froeb_cqg_2018,froeb_lima_cqg_2018}. The idea is to construct them as solutions of scalar differential equations on the perturbed spacetime that are trivially satisfied by the background coordinates. This method is particularly suited to the construction of invariant observables in perturbative gravity over highly symmetric backgrounds, such as the ones needed in cosmology. Moreover, it allows to easily construct the invariant observables at any order in perturbation theory. This method has already been successfully used in the computation of quantum gravitational loop corrections to invariant scalar correlators in Minkowski spacetime~\cite{froeb_cqg_2018} and of the quantum gravitational backreaction on the Hubble rate in inflationary cosmology~\cite{froeb_cqg_2019,lima_cqg_2021}. The method of Refs.~\cite{brunetti_etal_jhep_2016,froeb_cqg_2018,froeb_lima_cqg_2018} is an instance of the so-called geometrical clocks used in the relational approach, see for example Ref.~\cite{giesel_ijmpa_2008} and references therein; we believe it is the first instance where such clocks are explicitly constructed to all orders in perturbation theory.

Clearly, the choice of the field-dependent coordinate system is an integral part of the definition of the relational observable itself, and a specific reference frame might be more adequate to model the measurement at hand than others. In many applications in cosmology it is common place to express the line element of the FLRW spacetime in terms of conformally flat coordinates,
\begin{equation}
\label{eq:flrw_conformallyflat}
\total s^2 = a^2(\eta) \left( - \total\eta^2 + \total \vec{x}^2 \right) \eqend{,}
\end{equation}
where $a$ is the scale factor and $\eta$ is conformal time. This choice of coordinates makes the spacetime symmetries evident and simpler to exploit when decoupling and evolving the relevant degrees of freedom of the perturbations on top of this background~\cite{mukhanov_feldman_brandenberger_pr_1992}. However, it is not necessarily simple to construct relevant observables in these coordinates. This is the case for observables modeling measurements along the observer's past lightcone, which is how most of the experimental data in cosmology is actually collected. There are examples of coordinate systems adapted to measurements on the lightcone, such as the optical coordinates~\cite{temple_prsa_1938}, the observational coordinates~\cite{maartens_phdthesis_1980,ellis_etal_pr_1985,maartens_matravers_cqg_1994}, and coordinates based on the past lightcone of geodesic observers, such as the proposal of Ref.~\cite{preston_poisson_prd_2006} and the geodesic lightcone (GLC) coordinates~\cite{gasperini_et_al_jcap_2011}.

In this work we construct a new class of relational observables adapted to measurements on the observer's past lightcone and based on the GLC coordinates, using the approach of Refs.~\cite{brunetti_etal_jhep_2016,froeb_cqg_2018,froeb_lima_cqg_2018}. The GLC coordinates have the advantage of simplifying the computation of certain ``lightcone observables'', such as lightcone averages~\cite{gasperini_et_al_jcap_2011} and the luminosity-redshift relation~\cite{fanizza_etal_jcap_2015} --- for a review on the applications of the GLC coordinates, see Ref.~\cite{nugier_mg_2015}. The issue of constructing gauge-invariant observables using the GLC coordinates has been recently addressed in Refs.~\cite{fanizza_etal_jcap_2021,mitsou_etal_cqg_2021}, at first order in perturbation theory.\footnote{For related ideas in the context of black holes, see Refs.~\cite{nugier_jcap_2016, giddings_weinberg_prd_2020}.} The method we use allows us to more easily obtain explicit expressions for invariant observables beyond linear order. Moreover, the invariant observables we will define are somewhat different from the ones proposed in Ref.~\cite{fanizza_etal_jcap_2021}; we will discuss their exact relation later on in this paper. We will also address the time evolution of the observables at first order by deriving the linearised Einstein's equations when the spacetime is sourced by a scalar field (single-field inflation). Our derivation relies on the gauge-invariance of the linearised Einstein's equations when the background ones are satisfied. It differs from the one presented in Ref.~\cite{mitsou_etal_cqg_2021}, which makes use of ADM techniques.

The paper is organised as follows: In Sec.~\ref{sec:gauge_invariant_obs}, we briefly present the method of Refs.~\cite{brunetti_etal_jhep_2016,froeb_cqg_2018,froeb_lima_cqg_2018} to construct gauge-invariant observables in perturbative gravity. As an addition to it, here we introduce a gauge-invariant derivative operator that greatly simplifies some of the calculations we will face later. In Sec.~\ref{sec:perturbed_glc_coord}, we construct the field-dependent coordinates that enter the construction of our observables as non-linear GLC coordinates in a perturbed FLRW spacetime, and discuss their gauge freedom in the context of perturbative gravity. In Sec.~\ref{sec:gauge_invariant_glc_metric}, we analyse the invariant observable corresponding to the perturbed spacetime metric (which we call invariant perturbed metric in the following), up to second order in perturbation theory. In particular, there we show that the invariant perturbed metric coincides with the metric in the GLC gauge. In Sec.~\ref{sec:linearised_einstein_eqs}, we present our alternative derivation of the linearised Einstein's equations for a single-field inflationary model. We also compare the standard decomposition of the metric perturbation in the conformally flat coordinates with the one found in Sec.~\ref{sec:gauge_invariant_glc_metric}, and obtain an expression for the Sasaki--Mukhanov variable in the GLC coordinates. In Sec.~\ref{sec:hubble_parameter}, we study in detail two distinct definitions of the invariant Hubble rate (the invariant observable measuring the local spacetime expansion in single-field inflation), namely the one measured by geodesic observers and the one seen by observers co-moving with the inflaton. The discussion of our results and conclusions are presented in Sec.~\ref{sec:discussions}. We use the $+++$ convention of Ref.~\cite{misner_thorne_wheeler_book} in an $n$-dimensional spacetime, work in units such that $c = \hbar = 1$ and define $\kappa^2 \equiv 16 \pi G_\mathrm{N}$.

\section{Gauge-invariant observables}                                                                                                               %
\label{sec:gauge_invariant_obs}                                                                                                                     %

We consider a spacetime with coordinates $x^\mu$ and metric $\tilde{g}_{\mu\nu}$, and assume that the metric can be written as a background metric $g_{\mu\nu}$ and perturbations:
\begin{equation}
\label{eq:perturbed_metric}
\tilde{g}_{\mu\nu} \equiv g_{\mu\nu} + \kappa g_{\mu\nu}^{(1)} \eqend{.}
\end{equation}
Up to second order, the inverse perturbed metric is given by
\begin{equation}
\label{eq:inverse_perturbed_metric}
\tilde{g}^{\mu\nu} = g^{\mu\nu} - \kappa g^{\mu\nu}_{(1)} + \kappa^2 g_{\rho\sigma} g^{\mu\rho}_{(1)} g^{\sigma\nu}_{(1)} + \bigo{\kappa^3} \eqend{,}
\end{equation}
where the indices in the right-hand side of Eq.~(\ref{eq:inverse_perturbed_metric}) are raised and lowered with the background metric. Diffeomorphisms $x^\mu \to x^\mu - \kappa \xi^\mu(x)$ preserve the background metric, but imply the following gauge transformation for $g^{(1)}_{\mu\nu}$:
\begin{equation}
\delta_\xi g^{(1)}_{\mu\nu} = \lie_\xi \tilde{g}_{\mu\nu} \eqend{,}
\end{equation}
where $\lie_\xi$ denotes the Lie derivative with respect to the vector field $\xi^\mu$.

We would like to (perturbatively) construct a set of scalar fields $\tilde{X}^{(\mu)} = \tilde{X}^{(\mu)}[\tilde{g},x]$, with $\mu = 0,1,\ldots,n-1$, as functionals of the gauge degrees of freedom of the perturbed metric. [Since $\tilde{X}^{(\mu)}$ are scalars, we place the spacetime index $\mu$ within parentheses.] They will later serve as a coordinate frame, with respect to which we will measure the observables in the perturbed spacetime. A possible way to construct the fields $\tilde{X}^{(\mu)}[\tilde{g},x]$ is as solutions of some set of scalar differential equations, that we write as~\cite{brunetti_etal_jhep_2016}
\begin{equation}
\label{eq:X_tilde_eq}
D_{\tilde{g}}^{(\mu)}(\tilde{X}) = 0 \eqend{,}
\end{equation}
where $D_{\tilde{g}}^{(\mu)}$ are (possibly non-linear) differential operators involving the perturbed metric. Constructed in this way, the $\tilde{X}^{(\mu)}[\tilde{g},x]$ are in general non-local functionals of the metric perturbation. The form of the scalar equations~\eqref{eq:X_tilde_eq} is arbitrary in principle, but we require them to be (i) such that for the background $D_{g}^{(\mu)}(x) \equiv 0$, and (ii) causal, \ie, their solution should depend only on the perturbations within the past lightcone of the point $x$. Condition (i) guarantees that at the background level we have $\tilde{X}^{(\mu)}[g,x] = x^\mu$, while (ii) avoids unphysical effects coming from arbitrarily large spacelike separations in the invariant observables that are constructed with the help of these coordinates~\cite{froeb_lima_cqg_2018,froeb_cqg_2019}.

To illustrate this idea, let us consider perturbations in Minkowski spacetime with Cartesian coordinates. In this case, the coordinates $x^\mu$ satisfy, \eg, $\partial^\rho \partial_\rho x^\mu = 0$. Thus, a simple choice for the $\tilde{X}^{(\mu)}$ is to define them as solutions of the equation~\cite{froeb_cqg_2018}
\begin{equation}
\label{eq:harmonic_coordinates}
\tilde{\nabla}^\rho \tilde{\nabla}_\rho \tilde{X}^{(\mu)}[\tilde{g},x] = 0 \eqend{,}
\end{equation}
where $\tilde{\nabla}_\mu$ denotes the covariant derivative of the perturbed metric $\tilde{g}_{\mu\nu}$. To satisfy requirement (ii), one has to solve Eq.~\eqref{eq:harmonic_coordinates} using the Minkowski retarded Green's function at each order, which gives (with $\tilde{g}_{\mu\nu} = \eta_{\mu\nu} + \kappa h_{\mu\nu}$)
\begin{equation}
\label{eq:harmonic_coordinates_mink_firstorder}
\tilde{X}^{(\mu)}[\tilde{g},x] = x^\mu + \kappa \int G_\text{ret}(x,y) \left[ \partial_\nu h^{\mu\nu}(y) - \frac{1}{2} \partial^\mu h(y) \right] \total^n y + \bigo{\kappa^2} \eqend{.}
\end{equation}
In fact, the coordinates defined by Eq.~\eqref{eq:harmonic_coordinates} can be employed in perturbed spacetimes around arbitrary backgrounds, as long as one uses harmonic coordinates for the background.

Once the fields $\tilde{X}^{(\mu)}[\tilde{g},x]$ have been constructed, we can measure any tensor field with respect to them (and thus obtain a relational observable) by performing a field-dependent diffeomorphism from the background coordinates to the field-dependent ones. In particular, consider a tensor $\tilde{T}^{\alpha_1\dots \alpha_k}_{\beta_1\dots\beta_m}$ in the perturbed spacetime. Its transformation to the coordinates $\tilde{X}^{(\mu)}[\tilde{g},x]$ reads
\begin{equation}
\label{eq:gauge_invariant_tensor}
\mathcal{T}^{\mu_1\dots\mu_k}_{\nu_1\dots\nu_m}(\tilde{X}) = \frac{\partial \tilde{X}^{(\mu_1)}}{\partial x^{\alpha_1}}\dots \frac{\partial \tilde{X}^{(\mu_k)}}{\partial x^{\alpha_k}} \frac{\partial x^{\beta_1}}{\partial \tilde{X}^{(\nu_1)}}\dots \frac{\partial x^{\beta_m}}{\partial \tilde{X}^{(\nu_m)}} \tilde{T}^{\alpha_1\dots \alpha_k}_{\beta_1\dots\beta_m}[x(\tilde{X})]\eqend{,}
\end{equation}
where $x^\mu(\tilde{X})$ denotes the inverse of $\tilde{X}^{(\mu)}(x)$. The so-defined observable $\mathcal{T}^{\mu_1\dots\mu_k}_{\nu_1\dots\nu_m}$ is now invariant under (field-independent) diffeomorphisms of the spacetime manifold, \ie, while we have a non-trivial change
\begin{equation}
\delta_\xi \tilde{T}^{\alpha_1\dots \alpha_k}_{\beta_1\dots\beta_m}(x) = \lie_\xi \tilde{T}^{\alpha_1\dots \alpha_k}_{\beta_1\dots\beta_m}(x) \neq 0
\end{equation}
under the diffeomorphism $x^\mu \to x^\mu - \kappa \xi^\mu(x)$, for our invariant observable it results that
\begin{equation}
\delta_\xi \mathcal{T}^{\mu_1\dots\mu_k}_{\nu_1\dots\nu_m}(\tilde{X}) = 0 \eqend{,}
\end{equation}
since the transformation of the tensor components $\tilde{T}^{\alpha_1\dots \alpha_k}_{\beta_1\dots\beta_m}$ is compensated by the transformation of the $\tilde{X}^{(\mu)}[\tilde{g},x]$, which transform as scalar fields under diffeomorphisms since they fulfill a scalar equation.

As an useful example, let us consider Eq.~\eqref{eq:gauge_invariant_tensor} when our relational observable is a scalar $\tilde{S}$ in the perturbed spacetime. We first write the perturbative expansion of the field-dependent coordinates $\tilde{X}^{(\mu)}$ in the form
\begin{equation}
\label{eq:expansion_X_tilde}
\tilde{X}^{(\mu)}[\tilde{g},x] = x^\mu + \sum_{\ell = 1}^\infty \kappa^\ell X^{(\mu)}_{(\ell)}(x) \eqend{.}
\end{equation}
The perturbed scalar $\tilde{S}$ can be expanded as
\begin{equation}
\label{eq:expansion_S}
 \tilde{S}(x) = S(x) + \sum_{\ell = 1}^\infty \kappa^\ell S^{(\ell)}(x) \eqend{,}
\end{equation}
where $S$ is its background value. To obtain the perturbative expansion for $x^\mu(\tilde{X})$, we need to invert the relation~\eqref{eq:expansion_X_tilde}. This can be easily done up to second order in $\kappa$:
\begin{splitequation}
\label{eq:x_functional_X_tilde}
x^\mu(\tilde{X}) &= \tilde{X}^{(\mu)} - \kappa X^{(\mu)}_{(1)}(x) - \kappa^2 X^{(\mu)}_{(2)}(x) + \bigo{\kappa^3} \\
&= \tilde{X}^{(\mu)} - \kappa X^{(\mu)}_{(1)}(\tilde{X} - \kappa X_{(1)}) - \kappa^2 X^{(\mu)}_{(2)}(\tilde{X}) + \bigo{\kappa^3} \\
&= \tilde{X}^{(\mu)} - \kappa X^{(\mu)}_{(1)}(\tilde{X}) - \kappa^2 \left[ X^{(\mu)}_{(2)}(\tilde{X}) - X^{(\nu)}_{(1)}(\tilde{X}) \partial_\nu X^{(\mu)}_{(1)}(\tilde{X}) \right] + \bigo{\kappa^3} \eqend{.}
\end{splitequation}
(Higher orders are straightforward but lengthy.) Then, by combining Eqs.~\eqref{eq:expansion_S} and~\eqref{eq:x_functional_X_tilde}, we compute the gauge-invariant observable corresponding to $\tilde{S}$ as
\begin{splitequation}
\label{eq:scalar_observable_expansion}
\mathcal{S}(\tilde{X}) &\equiv \tilde{S}[x(\tilde{X})] = \sum_{\ell = 0}^\infty \kappa^\ell \mathcal{S}^{(\ell)}(\tilde{X}) \\
&= S(\tilde{X}) + \kappa \left[ S^{(1)}(\tilde{X}) - X^{(\sigma)}_{(1)}(\tilde{X}) \partial_\sigma S(\tilde{X}) \right] \\
&\quad+ \kappa^2 \bigg[ S^{(2)}(\tilde{X}) - X^{(\sigma)}_{(1)}(\tilde{X}) \partial_\sigma S^{(1)}(\tilde{X}) + \frac{1}{2} X^{(\rho)}_{(1)}(\tilde{X}) X^{(\sigma)}_{(1)}(\tilde{X}) \partial_\rho \partial_\sigma S(\tilde{X}) \\
&\quad+ X^{(\rho)}_{(1)}(\tilde{X}) \partial_\rho X^{(\sigma)}_{(1)}(\tilde{X}) \partial_\sigma S(\tilde{X}) - X^{(\sigma)}_{(2)}(\tilde{X}) \partial_\sigma S(\tilde{X}) \bigg] + \bigo{\kappa^3}\eqend{.}
\end{splitequation}
Note that now the coordinates covering the spacetime are $\tilde{X}^{(\mu)}$, and since coordinates are mere labels, we can denote them again by $x^\mu$. That is, the first orders in the perturbative expansion of our invariant observable read
\begin{equations}[eq:scalar_observable_expansion2]
\mathcal{S}^{(0)} &= S \eqend{,} \\
\mathcal{S}^{(1)} &= S^{(1)} - X^{(\sigma)}_{(1)} \partial_\sigma S \eqend{,} \\
\mathcal{S}^{(2)} &= S^{(2)} - X^{(\sigma)}_{(1)} \partial_\sigma S^{(1)} + \frac{1}{2} X^{(\rho)}_{(1)} X^{(\sigma)}_{(1)} \partial_\rho \partial_\sigma S + \left[ X^{(\rho)}_{(1)} \partial_\rho X^{(\sigma)}_{(1)} - X^{(\sigma)}_{(2)} \right] \partial_\sigma S \eqend{.}
\end{equations}

We can use Eq.~\eqref{eq:scalar_observable_expansion2} to check that $\mathcal{S}$ is invariant under diffeomorphisms that preserve the background, of the form $x^\mu \to x^\mu - \kappa \xi^\mu(x)$. The scalars $\tilde{X}^{(\mu)}$ and $\tilde{S}$ transform as
\begin{splitequation}
\label{eq:gauge_transformation_X}
\delta_\xi \tilde{X}^{(\mu)} &\equiv \kappa \lie_\xi \tilde{X}^{(\mu)} = \kappa \xi^\mu \partial_\mu \tilde{X}^{(\mu)} \\
&= \kappa \delta_\xi X^{(\mu)}_{(1)} + \kappa^2 \delta_\xi X^{(\mu)}_{(2)} + \bigo{\kappa^3} \\
&= \kappa \xi^\mu + \kappa^2 \xi^\nu \partial_\nu \tilde{X}^{(\mu)}_{(1)} + \bigo{\kappa^3}
\end{splitequation}
and
\begin{splitequation}
\label{eq:gauge_transformation_S}
\delta_\xi \tilde{S} & = \kappa \xi^\mu \partial_\mu \tilde{S} \\
&= \kappa \delta_\xi S^{(1)} + \kappa^2 \delta_\xi S^{(2)} + \bigo{\kappa^3} \\
&= \kappa \xi^\mu \partial_\mu S + \kappa^2 \xi^\mu \partial_\mu S^{(1)} + \bigo{\kappa^3} \eqend{.}
\end{splitequation}
To compute the transformation of Eq.~\eqref{eq:scalar_observable_expansion2}, we substitute Eqs.~\eqref{eq:gauge_transformation_X} and~\eqref{eq:gauge_transformation_S} into the expression for $\delta_\xi \mathcal{S}$. At first order, for example, we obtain
\begin{equation}
\delta_\xi \mathcal{S}^{(1)} = \delta_\xi S^{(1)} - \delta_\xi X^{(\mu)}_{(1)} \partial_\mu S = \xi^\mu \partial_\mu S - \xi^\mu \partial_\mu S = 0 \eqend{,}
\end{equation}
and similar results follow at higher orders in perturbation theory and for vector and tensor fields.

The perturbed metric tensor itself can be turned into an invariant observable. Indeed, using Eq.~\eqref{eq:gauge_invariant_tensor} we define the invariant perturbed metric as
\begin{equation}
\label{eq:inv_metric}
\mathcal{G}_{\mu\nu}(\tilde{X}) \equiv \frac{\partial x^{\rho}}{\partial \tilde{X}^{(\mu)}}\frac{\partial x^{\sigma}}{\partial \tilde{X}^{(\nu)}} \tilde{g}_{\rho\sigma}[x(\tilde{X})] \eqend{.}
\end{equation}
Using Eqs.~\eqref{eq:perturbed_metric} and~\eqref{eq:x_functional_X_tilde}, we can separate the expression above into its background and perturbation parts as
\begin{equation}
\label{eq:inv_metric_perturbation}
\mathcal{G}_{\mu\nu} = g_{\mu\nu} + \kappa \mathcal{H}_{\mu\nu} \eqend{,}
\end{equation}
where $\mathcal{H}_{\mu\nu}$ denotes the invariant metric perturbation. The invariant metric can be expanded as
\begin{equation}
\mathcal{G}_{\mu\nu} = g_{\mu\nu} + \sum_{\ell = 1}^\infty \kappa^\ell \mathcal{G}^{(\ell)}_{\mu\nu} \eqend{,}
\end{equation}
such that
\begin{equation}
\mathcal{H}_{\mu\nu} = \sum_{\ell = 0}^\infty \kappa^\ell \mathcal{G}^{(\ell + 1)}_{\mu\nu} \eqend{.}
\end{equation}
By combining Eqs.~\eqref{eq:perturbed_metric} and~\eqref{eq:x_functional_X_tilde}, we find that the first- and second-order terms of the perturbative series for $\mathcal{G}_{\mu\nu}$ are given by
\begin{equation}
\label{eq:G_1}
\mathcal{G}_{\mu\nu}^{(1)} = g_{\mu\nu}^{(1)} - X_{(1)}^{(\rho)} \partial_\rho g_{\mu\nu} - g_{\mu\rho} \partial_\nu X_{(1)}^{(\rho)} - g_{\rho\nu} \partial_\mu X_{(1)}^{(\rho)}
\end{equation}
and
\begin{splitequation}
\label{eq:G_2}
\mathcal{G}_{\mu\nu}^{(2)} &= - \left[ X^{(\rho)}_{(2)} - X^{(\sigma)}_{(1)} \partial_\sigma X^{(\rho)}_{(1)} \right] \partial_\rho g_{\mu\nu} + \frac{1}{2} X^{(\rho)}_{(1)} X^{(\sigma)}_{(1)} \partial_\rho \partial_\sigma g_{\mu\nu} + \partial_\mu X^{(\rho)}_{(1)} \partial_\nu X^{(\sigma)}_{(1)} g_{\rho\sigma} \\
&\quad- 2 \left[ g^{(1)}_{\rho(\mu} - X^{(\sigma)}_{(1)} \partial_\sigma g_{\rho(\mu} \right] \partial_{\nu)} X^{(\rho)}_{(1)} - 2 g_{\rho(\mu} \partial_{\nu)} \left[ X^{(\rho)}_{(2)} - X^{(\sigma)}_{(1)} \partial_\sigma X^{(\rho)}_{(1)} \right] - X^{(\sigma)}_{(1)} \partial_\sigma g_{\mu\nu}^{(1)} \eqend{.}
\end{splitequation}
For later use, we also define the invariant inverse metric $\mathcal{G}^{\mu\nu}$ as
\begin{equation}
\label{eq:inverse_inv_metric}
\mathcal{G}^{\mu\nu}(\tilde{X}) \equiv \frac{\partial \tilde{X}^{(\mu)}}{\partial x^{\rho}}\frac{\partial \tilde{X}^{(\nu)}}{\partial x^{\sigma}} \tilde{g}^{\rho\sigma}[x(\tilde{X})] \eqend{.}
\end{equation}
Its perturbative series can be written as
\begin{equation}
\mathcal{G}^{\mu\nu} = g^{\mu\nu} + \sum_{\ell = 1}^\infty \kappa^\ell \mathcal{G}^{\mu\nu}_{(\ell)} \eqend{,}
\end{equation}
and it is related to the invariant metric perturbation $\mathcal{H}_{\mu\nu}$ as
\begin{equation}
\label{eq:inverse_inv_metric_hmunu_expansion}
\mathcal{G}^{\mu\nu} = g^{\mu\nu} - \kappa \mathcal{H}^{\mu\nu} + \kappa^2 \mathcal{H}^{\mu\rho} \, \mathcal{H}_\rho{}^\nu + \bigo{\kappa^3} \eqend{.}
\end{equation}
Its first- and second-order terms read
\begin{equation}
\label{eq:inverse_G_1}
\mathcal{G}^{\mu\nu}_{(1)} = - g^{\mu\nu}_{(1)} - g^{\mu\sigma} \partial_\sigma X^{(\nu)}_{(1)} - g^{\nu\sigma} \partial_\sigma X^{(\mu)}_{(1)} + X^{(\sigma)}_{(1)} \partial_\sigma g^{\mu\nu}
\end{equation}
and
\begin{splitequation}
\label{eq:inverse_G_2}
\mathcal{G}_{(2)}^{\mu\nu} &= g_{\rho\sigma} g^{\mu\rho}_{(1)} g^{\sigma\nu}_{(1)} + g^{\mu\sigma}_{(1)} \partial_\sigma X^{(\nu)}_{(1)} + g^{\nu\sigma}_{(1)} \partial_\sigma X^{(\mu)}_{(1)} - X^{(\sigma)}_{(1)} \partial_\sigma g^{\mu\nu}_{(1)} - X^{(\sigma)}_{(2)} \partial_\sigma g^{\mu\nu} \\
&\quad+ g^{\mu\rho} \partial_\rho X^{(\nu)}_{(2)} + g^{\nu\rho} \partial_\rho X^{(\mu)}_{(2)} + X^{(\rho)}_{(1)} \partial_\rho X^{(\sigma)}_{(1)} \partial_\sigma g^{\mu\nu} - g^{\mu\rho} X^{(\sigma)}_{(1)} \partial_\rho \partial_\sigma X^{(\nu)}_{(1)} \\
&\quad- g^{\nu\rho} X^{(\sigma)}_{(1)} \partial_\rho \partial_\sigma X^{(\mu)}_{(1)} - X^{(\rho)}_{(1)} \partial_\rho g^{\mu\sigma} \partial_\sigma X^{(\nu)}_{(1)} - X^{(\rho)}_{(1)} \partial_\rho g^{\nu\sigma} \partial_\sigma X^{(\mu)}_{(1)} \\
&\quad+ g^{\rho\sigma} \partial_\rho X^{(\mu)}_{(1)} \partial_\sigma X^{(\nu)}_{(1)} + \frac{1}{2} X^{(\rho)}_{(1)} X^{(\sigma)}_{(1)} \partial_\rho \partial_\sigma g^{\mu\nu} \eqend{,}
\end{splitequation}
where we have used Eqs.~\eqref{eq:inverse_perturbed_metric} and~\eqref{eq:expansion_X_tilde}. Using these expressions, we can explicitly check, up to second order in perturbation theory, that $\mathcal{G}^{\mu\rho} \mathcal{G}_{\rho\nu} = \delta^\mu_\nu$. That is, our construction ensures that invariantisation and taking the inverse commute: the invariant inverse metric~\eqref{eq:inverse_inv_metric} is indeed the inverse of the invariant metric~\eqref{eq:inv_metric}.

Observables often also involve derivatives of the metric perturbation and potentially other tensor fields, such as in the example of the invariant Hubble rate that we will consider in Sec.~\ref{sec:hubble_parameter}. For these cases, it is useful to have a definition of an invariant covariant derivative, \ie, a covariant derivative associated with the metric $\mathcal{G}_{\mu\nu}$. Such a derivative operator can be easily found by recalling that the action of any two covariant derivative operators differ by a tensor field~\cite{wald_gr_book}. In particular, the covariant derivative of the invariant metric, here denoted by $\bnabla_\mu$, can be expressed in terms of the background covariant derivative $\nabla_\mu$ as
\begin{equation}
\label{eq:inv_derivative}
\bnabla_\mu \omega_\nu = \nabla_\mu \omega_\nu - \mathcal{C}^\rho_{\mu\nu} \omega_\rho \eqend{,}
\end{equation}
where the invariant tensor $\mathcal{C}^\rho_{\mu\nu}$ is given by
\begin{equation}
\label{eq:C}
\mathcal{C}^\rho_{\mu\nu} = \mathcal{G}^{\rho\sigma} \left( \nabla_{(\mu} \mathcal{G}_{\nu)\sigma} - \frac{1}{2} \nabla_\sigma \mathcal{G}_{\mu\nu} \right) \eqend{.}
\end{equation}
The invariant covariant derivative operator $\bnabla_\mu$ allows us to perform calculations directly in the field-dependent coordinate system $\tilde{X}^{(\mu)}$, rather than performing them first in the background coordinates and then transforming the result using Eq.~\eqref{eq:gauge_invariant_tensor} to produce a gauge-invariant expression. This will considerably shorten some of the computations presented in Sec.~\ref{sec:hubble_parameter}.

\section{Perturbed geodesic lightcone coordinates}                                                                                                  %
\label{sec:perturbed_glc_coord}                                                                                                                     %

\subsection{Basics on the geodesic lightcone coordinates}
\label{sec:basics_glc_coordinates}

We are interested in $n$-dimensional cosmological spacetimes in the so-called GLC coordinates~\cite{gasperini_et_al_jcap_2011}, whose construction we now recall. These coordinates are constructed by first picking a geodesic observer with proper time $\tau$. As long as this geodesic belongs to a congruence of irrotational timelike geodesics, it follows from Frobenius's theorem~\cite{wald_gr_book} that we can use the spatial section of this observer to foliate our spacetime with spatial hypersufaces labeled by $\tau$, whose normal is the four-velocity
\begin{equation}
\label{eq:u_mu}
u_\mu \equiv - \partial_\mu \tau \quad\text{with}\quad u^\mu u_\mu = -1 \eqend{.}
\end{equation}
To parametrise the spatial sections, we first foliate the spacetime with the past lightcones emanating from the observer's geodesic, and label these null hypersurfaces by $w$, with the null normal vector
\begin{equation}
\label{eq:k_mu}
k_\mu \equiv \partial_\mu w \quad\text{with}\quad k^\mu k_\mu = 0 \eqend{.}
\end{equation}
We remark that since the vectors fields $u^\mu$ and $k^\mu$ are defined as the normalised gradient of scalar fields, they are geodesic:
\begin{equation}
\label{eq:uk_geodesic}
u^\mu \nabla_\mu u^\rho = 0 = k^\mu \nabla_\mu k^\rho \eqend{.}
\end{equation}
Constructed in this way, the intersection of the hypersurfaces with fixed $\tau$ and $w$ is a spatial hypersurface isomorphic to the $(n-2)$-sphere, whose radius is equal to the time it takes the light to reach the observer. Assuming that the spacetime does not have caustics, \ie, that the observer's worldline is connected to any other spacetime point by a single null geodesic, any incoming light ray defines a fixed direction in the sky. Hence, one can parametrise the hypersurface corresponding to the celestial sphere with the angles $\theta^a$, with $a = 1, \ldots, n-2$, such that
\begin{equation}
\label{eq:k_mu_theta_a}
k^\mu \partial_\mu \theta^a = 0 \eqend{.}
\end{equation}

We denote by $(\partial_\tau)^\mu$, $(\partial_w)^\mu$ and $(\partial_a)^\mu$ the coordinate vectors defined by the coordinates $x^\mu = (\tau, w, \theta^a)$, respectively. From their definition as coordinate vectors, we have that
\begin
{equation}\label{eq:coord_vec}
(\partial_\tau)^\mu u_\mu = -1 \eqend{,} \quad (\partial_w)^\mu k_\mu = 1 \quad\text{and}\quad (\partial_b)^\mu \partial_\mu \theta^a = \delta^a_b \eqend{,}
\end{equation}
and that the other contractions of the coordinate vectors with $u_\mu$, $k_\mu$ and $\partial_\mu \theta^a$ vanish. To express the metric components in this basis, we need to express the coordinate vectors in terms of the vectors $k^\mu$ and $u^\mu$. The vector $k^\mu$ is tangent to a radial null geodesic, which implies that $k^w = k^a = 0$, while the frequency measured by the observer is given by
\begin{equation}
\label{eq:Upsilon}
k^\tau = - k^\mu u_\mu \equiv - \Upsilon^{-1} \eqend{.}
\end{equation}
In conclusion,
\begin{equation}
\label{eq:tau_coord_vec}
(\partial_\tau)^\mu = - \Upsilon k^\mu \eqend{,}
\end{equation}
which shows that the time coordinate vector is tangent to the backwards lightcone. Using Eqs.~\eqref{eq:coord_vec} and~\eqref{eq:Upsilon}, the four-velocity can be expanded in the coordinate basis as
\begin{equation}
u^\mu = (\partial_\tau)^\mu + \frac{1}{\Upsilon} \left[ (\partial_w)^\mu + U^a (\partial_a)^\mu \right] \eqend{,}
\end{equation}
where $U^a$ is defined by
\begin{equation}
\label{eq:U_a}
U^a \equiv \Upsilon u^\mu \partial_\mu \theta^a \eqend{.}
\end{equation}
It follows that
\begin{equation}
\label{eq:w_coord_vec}
(\partial_w)^\mu = \Upsilon^2 k^\mu + \Upsilon u^\mu - U^a (\partial_a)^\mu \eqend{.}
\end{equation}
The coordinate vector $(\partial_w)^\mu$ defines a notion of radial direction via successive past lightcones emanating from the observer. This radial direction, however, does not necessarily coincide with the spatial direction of propagation of the null geodesics reaching the observer. Indeed, employing a $3+1$ decomposition of the null vector $k^\mu$, we have~\cite{fleury_nugier_fanizza_jcap_2016}
\begin{equation}
k^\mu = - \Upsilon^{-1} ( d^\mu + u^\mu ) \eqend{,}
\end{equation}
where $d^\mu$ is a normalised spatial vector, \ie, $d^\mu u_\mu = 0$ and $d^\mu d_\mu = 1$. Substituting this expression in Eq.~\eqref{eq:w_coord_vec}, we obtain
\begin{equation}
(\partial_w)^\mu = - \Upsilon d^\mu - U^a (\partial_a)^\mu \eqend{.}
\end{equation}
Hence, the scalars $U^a$ measure the failure of the light rays to propagate radially towards the observer due to light-bending effects and the rotation of the observer's local frame.

We now want to express the metric in terms of the scalars $\Upsilon$, $U^a$ and the induced metric on the celestial sphere, which we denote by $\gamma_{ab}$. By using Eqs.~\eqref{eq:coord_vec}, \eqref{eq:tau_coord_vec} and~\eqref{eq:w_coord_vec}, we can compute the metric components as follows:
\begin{equations}
g_{\tau\tau} &= g_{\mu\nu} (\partial_\tau)^\mu (\partial_\tau)^\nu = \Upsilon^2 k^\mu k_\mu = 0 \eqend{,} \\
g_{ab} &= g_{\mu\nu} (\partial_a)^\mu (\partial_b)^\nu \equiv \gamma_{ab} \eqend{,} \label{eq:celestial_sphere_metric}\\
g_{ww} &= g_{\mu\nu} (\partial_w)^\mu (\partial_w)^\nu = \left[ \Upsilon^2 k^\mu + \Upsilon u^\mu - U^a (\partial_a)^\mu \right] \left[ \Upsilon^2 k_\mu + \Upsilon u_\mu - U^b (\partial_b)_\mu \right] = \Upsilon^2 + U^a U_a \eqend{,}\\
g_{\tau w} &= g_{\mu\nu} (\partial_\tau)^\mu (\partial_w)^\nu = - \Upsilon k_\mu (\partial_w)^\mu = - \Upsilon\eqend{,} \\
g_{\tau a} &= g_{\mu\nu} (\partial_a)^\mu (\partial_a)^\nu = - \Upsilon k_\mu (\partial_a)^\mu = 0 \eqend{,} \\
g_{w a} &= g_{\mu\nu} (\partial_w)^\mu (\partial_a)^\nu = (\partial_a)^\mu \left[ \Upsilon^2 k_\mu + \Upsilon u_\mu - U^b (\partial_b)_\mu \right] = - U_a \eqend{,}
\end{equations}
where we have defined $U_a \equiv \gamma_{ab} U^b$. This corresponds to the line element in the GLC gauge~\cite{gasperini_et_al_jcap_2011}:
\begin{equation}
\label{eq:glc_metric}
\total s^2 = g_{\mu\nu} \total x^\mu \total x^\nu = - 2 \Upsilon \total\tau \total w + \left( \Upsilon^2 + U^a U_a \right) \total w^2 + \gamma_{ab} \Big( \total\theta^a - U^a \total w \Big)\Big( \total\theta^b - U^b \total w \Big) \eqend{.}
\end{equation}
In matrix form, the metric $g_{\mu\nu}$ and its inverse $g^{\mu\nu}$ read
\begin{equation}
\label{eq:glc_metric_matrix_form}
g_{\mu\nu} = \begin{pmatrix} 0 & - \Upsilon & 0 \\ - \Upsilon & \Upsilon^2 + U^a U_a & - U_b \\ 0 & - U_a & \gamma_{ab} \end{pmatrix} \quad\text{and}\quad
g^{\mu\nu} = \begin{pmatrix} - 1 & - \Upsilon^{-1} & - \Upsilon^{-1} U^b \\ - \Upsilon^{-1} & 0 & 0 \\ - \Upsilon^{-1} U^a & 0 & \gamma^{ab} \end{pmatrix} \eqend{,}
\end{equation}
respectively. For more details on the geometrical construction behind the GLC metric and its properties, see for example Refs.~\cite{nugier_phdthesis_2013,fleury_nugier_fanizza_jcap_2016}.

In general spacetimes, we expect that the GLC coordinates only cover a certain spacetime patch around the observer. As mentioned earlier in this section, the formation of caustics that happens when strong lensing effects are present would make the coordinates not injective in the whole spacetime and, thus, ill-defined at the caustic and beyond (as seen by the observer). Furthermore, a superluminal expansion phase in the observer's past would also restrict the usage of lightcone coordinates to the causal patch that can be seen by the observer. Our main interest, however, lies in cosmological settings at scales where deviations from the cosmological homogeneous background can be model by small inhomogeneities, and in this setting caustics do not appear. Moreover, since we are modeling measurements performed by a localised observer, a superluminal expansion phase in the past is unproblematic. In fact, the use of GLC coordinates seems advantageous in this regard, since it ensures that causality is correctly taken into account for the effects that may influence the observer.

\subsection{GLC coordinates on perturbed cosmological spacetimes}

Let us now consider perturbations around an $n$-dimensional, spatially flat FLRW spacetime background with metric $g_{\mu\nu}$. The conformally flat coordinates yield the line element
\begin{equation}
\label{eq:flrw_metric_conformal}
\total s^2 = a^2(\eta) \left( - \total\eta^2 + \total r^2 + r^2 s_{ab} \total\theta^a \total\theta^b \right) \eqend{,}
\end{equation}
where $s_{ab}$ denotes the metric on $\mathbb{S}^{n-2}$, the unit $(n-2)$-sphere. On top of this homogeneous and isotropic background, we add perturbations to the metric according to
\begin{equation}
\label{eq:perturbed_metric_flrw}
\tilde{g}_{\mu\nu} = g_{\mu\nu} + \kappa a^2 h_{\mu\nu} \eqend{,}
\end{equation}
and the full inverse metric $\tilde{g}^{\mu\nu}$ reads
\begin{equation}
\label{eq:inverse_perturbed_metric_flrw}
\tilde{g}^{\mu\nu} = g^{\mu\nu} - \kappa a^2 h^{\mu\nu} + \kappa^2 a^4 h^{\mu\rho} h_\rho{}^\nu + \bigo{\kappa^3} \eqend{.}
\end{equation}
We wish to construct the GLC coordinates with the perturbed spacetime metric $\tilde{g}_{\mu\nu}$, up to second order in the perturbation $h_{\mu\nu}$. We again take a foliation for the full spacetime using an observer with proper time $\tilde{\tau}$, whose normal is the four-velocity
\begin{equation}
\label{eq:u_mu_tilde}
\tilde{u}_\mu \equiv - \partial_\mu \tilde{\tau} \quad\text{with}\quad \tilde{u}^\mu \tilde{u}_\mu = -1 \eqend{,}
\end{equation}
and expand the proper time and $\tilde{u}_\mu$ as
\begin{equation}
\label{eq:tau_tilde_expansion}
\tilde{\tau} = \tau + \kappa \tau_{(1)} + \kappa_2 \tau_{(2)} + \bigo{\kappa^3}
\end{equation}
and
\begin{splitequation}
\label{eq:u_mu_tilde_expansion}
\tilde{u}_\mu &= u_\mu + \kappa u_\mu^{(1)} + \kappa^2 u_\mu^{(2)} + \bigo{\kappa^3} \\
&= - \delta_\mu^\tau - \kappa \partial_\mu \tau_{(1)} - \kappa^2 \partial_\mu \tau_{(2)} + \bigo{\kappa^3} \eqend{.}
\end{splitequation}
The spatial section is parametrised using the observer's past lightcones labeled by the parameter $\tilde{w}$, and $n-2$ angular coordinates $\tilde{\theta}^a$. The corresponding null vector is given by
\begin{equation}
\label{eq:k_mu_tilde}
\tilde{k}_\mu \equiv \partial_\mu \tilde{w} \quad\text{with}\quad \tilde{k}^\mu \tilde{k}_\mu = 0 \eqend{,}
\end{equation}
and the angular coordinates are defined such that
\begin{equation}
\label{eq:theta_tilde}
\tilde{k}^\mu \partial_\mu \tilde{\theta}^a = 0 \eqend{.}
\end{equation}
The expansion of the lightcone coordinate reads
\begin{equation}
\label{eq:w_tilde_expansion}
\tilde{w} = w + \kappa w_{(1)} + \kappa^2 w_{(2)} + \bigo{\kappa^3}
\end{equation}
and the full null vector is expanded as
\begin{splitequation}
\label{eq:k_mu_tilde_expansion}
\tilde{k}_\mu &= k_\mu + \kappa k_\mu^{(1)} + \kappa^2 k_\mu^{(2)} + \bigo{\kappa^3} \\
&= \delta_\mu^w + \kappa \partial_\mu w_{(1)} + \kappa^2 \partial_\mu w_{(2)} + \bigo{\kappa^3} \eqend{,}
\end{splitequation}
while the expansion for the angular coordinates reads
\begin{equation}
\label{eq:theta_tilde_expansion}
\tilde{\theta}^a = \theta^a + \kappa \theta_{(1)}^a + \kappa^2 \theta_{(2)}^a + \bigo{\kappa^3} \eqend{.}
\end{equation}
Equations~\eqref{eq:u_mu_tilde}, \eqref{eq:k_mu_tilde} and~\eqref{eq:theta_tilde} are the scalar equations defining our field-dependent GLC coordinates $\tilde{X}^{(\mu)} = (\tilde{\tau},\tilde{w}, \tilde{\theta}^a)$ discussed in Sec.~\ref{sec:gauge_invariant_obs}, and will be used later on to define a class of gauge-invariant lightcone observables in inflationary spacetimes.

It is not difficult to cast the FLRW metric in conformally flat coordinates into the form of Eq.~\eqref{eq:glc_metric}. Indeed, by using the coordinate
\begin{equation}
\label{eq:w_coord_flrw}
w \equiv r + \eta(\tau)
\end{equation}
in Eq.~\eqref{eq:flrw_metric_conformal}, we obtain the background metric $g_{\mu\nu}$ in the GLC coordinates $x^\mu = (\tau, w, \theta^a)$:
\begin{equation}
\label{eq:flrw_metric_glc}
\total s^2 = g_{\mu\nu} \total x^\mu \total x^\nu = - 2 a \total\tau \total w + a^2 \total w^2 + a^2 (w - \eta)^2 s_{ab} \total\theta^a \total\theta^b \eqend{,}
\end{equation}
where the background observer's proper time $\tau$ is obtained from the conformal time $\eta$ by integrating the relation $\total\tau = a(\eta) \total\eta$. We thus have the identifications
\begin{equation}
\Upsilon = a \eqend{,} \quad U^a = 0 \quad\text{and}\quad \gamma_{ab} = a^2 (w - \eta)^2 s_{ab} \eqend{.}
\end{equation}
It is also straightforward to write down the expressions for the observer's four-velocity $u^\mu$ and the null vector $k^\mu$ in the FLRW spacetime. They are
\begin{equation}
\label{eq:u_mu_0}
u_\mu = (-1,0,0^a) \quad\text{and}\quad u^\mu = (1,a^{-1},0^a) \eqend{,}
\end{equation}
and
\begin{equation}
\label{eq:k_mu_0}
k_\mu = (0,1,0^a) \quad\text{and}\quad k^\mu = (- a^{-1},0,0^a) \eqend{.}
\end{equation}
We can check that both $u^\mu$ and $k^\mu$ satisfy the geodesic equation~\eqref{eq:uk_geodesic} and note that Eq.~\eqref{eq:k_mu_theta_a} is clearly satisfied. For later use, we also compute the Christoffel symbols for the FLRW metric in the GLC coordinates~\eqref{eq:flrw_metric_glc}. They are
\begin{equations}[eq:christoffel_symb]
\Gamma^\tau_{\mu\nu} &= H ( \delta_\mu^\tau \delta_\nu^\tau + g_{\mu\nu} ) \eqend{,} \\
\Gamma^w_{\mu\nu} &= H a \delta_\mu^w \delta_\nu^w + \frac{1}{a} \left[ H - \frac{1}{a(w - \eta)} \right] \gamma_{\mu\nu} \eqend{,} \\
\Gamma^a_{\mu\nu} &= 2 \left[ H - \frac{1}{a(w - \eta)} \right] \delta_{(\mu}^\tau \delta_{\nu)}^a + \frac{2}{w - \eta} \delta_{(\mu}^w \delta_{\nu)}^a + \gamma^{ab} \partial_c \gamma_{b(\mu} \delta_{\nu)}^c - \frac{1}{2} \gamma^{ab} \partial_b \gamma_{\mu\nu} \eqend{,}
\end{equations}
where $\gamma_{\mu\nu} \equiv \gamma_{ab} \delta^a_\mu \delta^b_\nu$. At this point, it is convenient to define
\begin{equation}
\label{eq:slow_roll_parameters}
H \equiv \frac{\dot a}{a} \eqend{,} \quad \epsilon \equiv - \frac{\dot H}{H^2} \quad\text{and}\quad \delta \equiv \frac{\dot\epsilon}{2 H \epsilon} \eqend{,}
\end{equation}
which are the Hubble parameter and the first and second slow-roll parameters of the background, respectively.\footnote{The slow-roll parameters defined in Eq.~\eqref{eq:slow_roll_parameters} are related to the widely used Hubble slow-roll parameters $\epsilon_H$ and $\eta_H$ as $\epsilon = \epsilon_H$ and $\delta = \epsilon - \eta_H$, see, \eg, Ref.~\cite{liddle_parsons_barrow_prd_1994}.} (The dot denotes the derivative with respect to $\tau$.)

To find the higher-order corrections to the GLC coordinates engendered by the perturbation in the full spacetime, it is enough to use the normalisation conditions~\eqref{eq:u_mu_tilde} and~\eqref{eq:k_mu_tilde} for $\tilde{u}^\mu$ and $\tilde{k}^\mu$, their expansions~\eqref{eq:u_mu_tilde_expansion} and~\eqref{eq:k_mu_tilde_expansion}, and the condition~\eqref{eq:theta_tilde} for the angular coordinates. We start with the temporal coordinate $\tilde{\tau}$. The expansion of Eq.~\eqref{eq:u_mu_tilde} up to second order in perturbation theory yields
\begin{splitequation}
-1 = \tilde{u}^\mu \tilde{u}_\mu &= - 1 + \kappa \left( 2 u^\mu u_\mu^{(1)} - a^2 h^{\mu\nu} u_\mu u_\nu \right) \\
&\quad+ \kappa^2 \left( - 2 u^\mu u^{(2)}_\mu + u^{\mu}_{(1)} u^{(1)}_\mu - 2 a^2 h^{\mu\nu} u_\mu u^{(1)}_\nu + a^4 h^{\mu\rho} h_\rho{}^\nu u_\mu u_\nu \right) + \bigo{\kappa^3} \eqend{.}
\end{splitequation}
Hence, using the expansion~\eqref{eq:u_mu_tilde_expansion}, the first- and second-order corrections to the time coordinate satisfy
\begin{equations}[eq:u_mu_1_2]
u^\mu \partial_\mu \tau^{(1)} &= - \frac{a^2}{2} u^\mu u^\nu h_{\mu\nu} \eqend{,} \\
u^\mu \partial_\mu \tau^{(2)} &= \frac{1}{2} \left( u^{\mu}_{(1)} u^{(1)}_\mu - 2 a^2 h^{\mu\nu} u_\mu u^{(1)}_\nu + a^4 h^{\mu\rho} h_\rho{}^\nu u_\mu u_\nu \right) \eqend{.}
\end{equations}
The corrections to the lightcone coordinate $\tilde{w}$ are obtained by computing
\begin{splitequation}
0 = \tilde{k}^\mu \tilde{k}_\mu &= \kappa \left( 2 k^\mu k_\mu^{(1)} - a^2 h^{\mu\nu} k_\mu k_\nu \right) \\
&\quad+ \kappa^2 \left( 2 k^\mu k_\mu^{(2)} + k^{\mu}_{(1)} k^{(1)}_\mu - 2 a^2 h^{\mu\nu} k_\mu k^{(1)}_\nu + a^4 h^{\mu\rho} h_\rho{}^\nu k_\mu k_\nu \right) + \bigo{\kappa^3} \eqend{,}
\end{splitequation}
which together with the expansion~\eqref{eq:k_mu_tilde_expansion} gives
\begin{equations}[eq:k_mu_1_2]
k^\mu \partial_\mu w^{(1)} & = \frac{a^2}{2} k^\mu k^\nu h_{\mu\nu} \eqend{,} \\
k^\mu \partial_\mu w^{(2)} & = - \frac{1}{2} \left( k^\mu_{(1)} k^{(1)}_\mu - 2 a^2 h^{\mu\nu} k_\mu k^{(1)}_\nu + a^4 h^{\mu\rho} h_\rho{}^\nu k_\mu k_\nu \right) \eqend{.}
\end{equations}
Finally, we use Eq.~\eqref{eq:theta_tilde} to obtain the equations that determine the corrections to the angular coordinates, and up to second order obtain
\begin{splitequation}
0 &= \tilde{k}^\mu \partial_\mu \tilde{\theta}^a = \kappa \left( k^\mu \partial_\mu \theta_{(1)}^a + k_{(1)}^a - a^2 h^{\mu a} k_\mu \right) \\
&\quad+ \kappa^2 \bigg( k^\mu \partial_\mu \theta_{(2)}^a + k_{(2)}^a + k^\mu_{(1)} \partial_\mu \theta^a_{(1)} - a^2 h^{\mu\nu} k_\mu \partial_\nu \theta^a_{(1)} - a^2 h^{\mu a} k^{(1)}_\mu + a^4 h^{\mu\rho} h_\rho{}^a k_\mu \bigg) + \bigo{\kappa^3} \eqend{,}
\end{splitequation}
which gives
\begin{equations}[eq:del_theta_mu_1_2]
k^\mu \partial_\mu \theta_{(1)}^a &= - k^a_{(1)} + a^2 h^{\mu a} k_\mu \eqend{,} \\
k^\mu \partial_\mu \theta_{(2)}^a &= - k^a_{(2)} - k^\mu_{(1)} \partial_\mu \theta^a_{(1)} + a^2 h^{\mu\nu} k_\mu \partial_\nu \theta^a_{(1)} + a^2 h^{\mu a} k^{(1)}_\mu - a^4 h^{\mu\rho} h_\rho{}^a k_\mu \eqend{.}
\end{equations}

To obtain more explicit expressions for Eqs.~\eqref{eq:u_mu_1_2}, \eqref{eq:k_mu_1_2} and~\eqref{eq:del_theta_mu_1_2}, we employ the basis defined by the background GLC coordinates. Hence, we use the background vectors~\eqref{eq:u_mu_0} and~\eqref{eq:k_mu_0} to obtain the following differential equations for the first-order corrections:
\begin{equations}[eq:eom_1]
\left[ \partial_\tau + a^{-1}(\eta) \partial_w \right] \tau_{(1)} &= - \frac{1}{2} \left[ a^2(\eta) h_{\tau\tau} + 2 a(\eta) h_{\tau w} + h_{ww} \right] \eqend{,} \label{eq:eom_1_tau} \\
\partial_\tau w_{(1)} &= - \frac{1}{2} a(\eta) h_{\tau\tau} \eqend{,} \label{eq:eom_1_w} \\
\partial_\tau \theta_{(1)}^a &= a(\eta) \gamma^{ab} \left[ \partial_b w_{(1)} + a(\eta) h_{\tau b} \right] \label{eq:eom_1_a} \eqend{.}
\end{equations}
These equations can be integrated after the initial conditions for the perturbed coordinates have been specified. We will assume that the metric perturbations are either localised (of compact support), or fall off fast enough for large $r = w - \eta$ and large negative $\tau$ (spatial and past infinity), and that there the perturbed coordinates coincide with the background ones. As a result, we obtain the convergent integral expressions
\begin{equations}[eq:X_1_integral]
\tau_{(1)}(\tau,w,\theta) &= - \frac{1}{2} \int^\tau_{-\infty} \left( a^2 h_{\tau\tau} + 2 a h_{\tau w} + h_{ww} \right)[s,w - \eta(\tau) + \eta(s),\theta] \total s \eqend{,} \\
w_{(1)}(\tau,w,\theta) &= - \frac{1}{2} \int^\tau_{-\infty} \left( a h_{\tau\tau} \right)(s,w,\theta) \total s \eqend{,} \\
\theta_{(1)}^a(\tau,w,\theta) &= \int^\tau_{-\infty} \left( a \gamma^{ab} \partial_b w_{(1)} + a^2 h^a{}_\tau \right)(s,w,\theta) \total s \eqend{.}
\end{equations}
These expressions for the corrections to the GLC coordinates show that they are non-local functionals of the metric perturbation. Namely, the integrals depend on the metric perturbation along the whole light ray reaching the observation point from past null infinity. Nevertheless, albeit non-local, these corrections depend only on the metric perturbation on the past lightcone emanating from the observation point, and are therefore causal.

The components of the first-order corrections to $\tilde{u}^\mu$ and $\tilde{k}^\mu$ can be easily worked out from Eq.~\eqref{eq:X_1_integral}, and read
\begin{equations}[eq:u_1_mu]
u^\tau_{(1)} &= \frac{1}{2} \left( a^2 h_{\tau\tau} + 2 a h_{\tau w} + h_{ww} \right) \eqend{,} \\
u^w_{(1)} &= a^{-1} \partial_\tau \tau_{(1)} + a h_{\tau\tau} + h_{\tau w} \eqend{,} \\
u^a_{(1)} &= - (w - \eta) \gamma^{ab} \left( \partial_b \tau_{(1)} - a^2 h_{\tau b} + a h_{w b} \right) \eqend{,}
\end{equations}
as well as
\begin{equations}[eq:k_1_mu]
k^\tau_{(1)} &= - a^{-1} \partial_w w_{(1)} - \frac{a}{2} h_{\tau\tau} - h_{\tau w} \eqend{,} \\
k^w_{(1)} & = - \frac{1}{2} h_{\tau\tau} \eqend{,} \\
k^a_{(1)} & = (w - \eta) \gamma^{ab} \left( \partial_b w_{(1)} + a h_{\tau b} \right) \eqend{.}
\end{equations}
Equations~\eqref{eq:u_1_mu} and~\eqref{eq:k_1_mu} are useful in the computation of the second-order corrections to the GLC coordinates.

The equations for the second-order corrections to the GLC coordinates are found by following the same steps we took at first order. By substituting Eqs.~\eqref{eq:u_mu_0} and~\eqref{eq:k_mu_0} and Eqs.~\eqref{eq:u_1_mu} and~\eqref{eq:k_1_mu} into Eqs.~\eqref{eq:u_mu_1_2}, \eqref{eq:k_mu_1_2} and~\eqref{eq:del_theta_mu_1_2}, we can write the equations for the second-order corrections in the form
\begin{equations}[eq:eom_2]
\begin{split}
\left( \partial_\tau + a^{-1} \partial_w \right) \tau_{(2)} &= \frac{1}{2} g^{\mu\nu} \partial_\mu \tau_{(1)} \partial_\nu \tau_{(1)} + \left( a^2 h_\tau{}^\mu + a h_w{}^\mu \right) \partial_\mu \tau_{(1)} \\
&\quad+ \frac{a^2}{2} \left( a^2 h_\tau{}^\mu h_{\mu\tau} + 2 a h_\tau{}^\mu h_{\mu w} + h_w{}^\mu h_{\mu w} \right) \eqend{,}
\end{split} \\
\partial_\tau w_{(2)} &= \frac{a}{2} \left( g^{\mu\nu} \partial_\mu w_{(1)} \partial_\nu w_{(1)} + 2 a h_\tau{}^\mu \partial_\mu w_{(1)} + a^2 h_\tau{}^\mu h_{\mu\tau} \right) \eqend{,} \\
\partial_\tau \theta_{(2)}^a &= a \gamma^{ab} \partial_b w_{(2)} + a g^{\mu\nu} \partial_\mu w_{(1)} \partial_\nu \theta_{(1)}^a + a^2 h_\tau{}^\mu \partial_\mu \theta_{(1)}^a - a^3 h^{a\mu} \left( \partial_\mu w_{(1)} + a h_{\mu\tau} \right) \eqend{.}
\end{equations}
Integrating these equations with the same boundary conditions as at first order, we obtain
\begin{equations}[eq:X_2_integral]
\begin{split}
\tau_{(2)}(\tau,w,\theta) &= \frac{1}{2}\int_{-\infty}^\tau \Big[ g^{\mu\nu} \partial_\mu \tau_{(1)} \partial_\nu \tau_{(1)} + 2 \left( a^2 h_\tau{}^\mu + a h_w{}^\mu \right) \partial_\mu \tau_{(1)} \\
&\qquad\qquad\quad+ a^4 h_\tau{}^\mu h_{\mu\tau} + 2 a^3 h_\tau{}^\mu h_{\mu w} + a^2 h_w{}^\mu h_{\mu w} \Big][s,w - \eta(\tau) + \eta(s),\theta] \total s \eqend{,}
\end{split} \\
w_{(2)}(\tau,w,\theta) &= \frac{1}{2} \int_{-\infty}^\tau \left( a g^{\mu\nu} \partial_\mu w_{(1)} \partial_\nu w_{(1)} + 2 a^2 h_\tau{}^\mu \partial_\mu w_{(1)} + a^3 h_\tau{}^\mu h_{\mu\tau} \right)(s,w,\theta) \total s \eqend{,} \\
\begin{split}
\theta_{(2)}^a(\tau,w,\theta) &= \int_{-\infty}^\tau \Big[ a \gamma^{ab} \partial_b w_{(2)} + a g^{\mu\nu} \partial_\mu w_{(1)} \partial_\nu \theta_{(1)}^a + a^2 h_\tau{}^\mu \partial_\mu \theta_{(1)}^a \\
&\qquad\qquad- a^3 h^{a\mu} \left( \partial_\mu w_{(1)} + a h_{\mu\tau} \right) \Big](s,w,\theta) \total s \eqend{.}
\end{split}
\end{equations}
Again, the second-order corrections are causal, non-local functionals of the metric perturbation $h_{\mu\nu}$.

With the explicit expressions for the corrections to the perturbed GLC coordinates, Eqs.~\eqref{eq:X_1_integral} and~\eqref{eq:X_2_integral}, we can check that $\tilde{X}^{(\mu)} = (\tilde{\tau},\tilde{w},\tilde{\theta}^a)$ transform as scalars under gauge transformations that leave the background invariant. This is a key aspect in constructing gauge-invariant observables using the method discussed in Sec.~\ref{sec:gauge_invariant_obs}. Thus, we again consider diffeomorphisms of the form $x^\mu \to x^\mu - \kappa \xi^\mu(x)$. Under this transformation, the metric perturbation transforms as
\begin{equation}
\label{eq:h_mu_nu_gauge_transformation}
a^2 \delta_\xi h_{\mu\nu} = \nabla_\mu \xi_\nu + \nabla_\nu \xi_\mu + \kappa \lie_\xi \left( a^2 h_{\mu\nu} \right) \eqend{,}
\end{equation}
where $\nabla_\mu$ is the covariant derivative of the FLRW background metric in the GLC coordinates~\eqref{eq:flrw_metric_glc}, with Christoffel symbols~\eqref{eq:christoffel_symb}, and we assume the same compact support or fall-off conditions on $\xi^\mu$ as for the metric perturbation $h_{\mu\nu}$.

We start by transforming the first-order corrections found in Eq.~\eqref{eq:X_1_integral}. For simplicity, let us consider the change in $w_{(1)}$ first. We compute to first order
\begin{splitequation}
\delta_\xi w_{(1)} &= - \frac{1}{2} \int^\tau_{-\infty} a[\eta(s)] \delta_\xi h_{\tau\tau}(s,w,\theta) \total s = - \int^\tau_{-\infty} \frac{1}{a[\eta(s)]} \left( \partial_s \xi_\tau - H \xi_\tau \right)(s,w,\theta) \total s \\
&= \int^\tau_{-\infty} \partial_s \xi^w(s,w,\theta) \total s = \xi^w(\tau,w,\theta) - \lim_{s \to -\infty} \xi^w(s,w,\theta) = \xi^w(\tau,w,\theta) \eqend{,}
\end{splitequation}
where we have used Eqs.~\eqref{eq:h_mu_nu_gauge_transformation} and~\eqref{eq:christoffel_symb}, and that by our support/fall-off conditions $\lim_{s \to -\infty} \xi^\mu(s,w,\theta) = 0$. The change in the time and angular coordinates lead to similar expressions, and the final result can be cast in the form
\begin{equation}
\label{eq:X_1_gauge_transformation}
\delta_\xi \tau_{(1)} = \xi^\tau = \xi^\mu \partial_\mu \tau \eqend{,} \qquad \delta_\xi w_{(1)} = \xi^w = \xi^\mu \partial_\mu w \eqend{,} \qquad \delta_\xi \theta_{(1)}^a = \xi^a = \xi^\mu \partial_\mu \theta^a \eqend{,}
\end{equation}
such that they do indeed transform as scalars to first order.

We now compute the changes in the second-order corrections to the GLC coordinates under a diffeomorphism. Notice that, since Eq.~\eqref{eq:h_mu_nu_gauge_transformation} has higher-order terms in $\kappa$, the transformation of the first-order corrections will now also produce second-order contributions. Using Eq.~\eqref{eq:h_mu_nu_gauge_transformation}, we find that including second-order contributions from the transformation of Eq.~\eqref{eq:X_1_integral} it follows that
\begin{equations}
\delta_\xi\tau_{(1)}(\tau,w,\theta) &= \xi^\tau - \frac{\kappa}{2} \int_{-\infty}^\tau \left[ \lie_\xi \left( a^2 h_{\mu\nu} \right) u^\mu u^\nu\right] [s, w - \eta(\tau) + \eta(s), \theta] \total s \eqend{,} \\
\delta_\xi w_{(1)}(\tau,w,\theta) &= \xi^w - \frac{\kappa}{2} \int_{-\infty}^\tau \left[ a \lie_\xi \left( a^2 h_{\mu\nu} \right) k^\mu k^\nu \right](s,w,\theta) \total s \eqend{,} \\
\delta_\xi\theta_{(1)}^a(\tau,w,\theta) &= \xi^a - \kappa \int_{-\infty}^\tau \left[ a \gamma^{ab} \lie_\xi \left( a^2 h_{\mu b} \right) k^\mu \right](s,w,\theta) \total s \eqend{.}
\end{equations}
Next, we apply the transformation to the second-order corrections in Eq.~\eqref{eq:X_2_integral}. For simplicity, let us consider the change in the correction $w_{(2)}$, which yields
\begin{splitequation}
\delta_\xi w_{(2)} &= \int_{-\infty}^\tau \Big[ a g^{\mu\nu} \partial_\mu w_{(1)} \partial_\nu \delta_\xi w_{(1)} + a^2 \delta_\xi h_{\tau\mu} g^{\mu\nu} \partial_\nu w_{(1)} + a^2 h_\tau{}^\mu \partial_\mu \delta_\xi w_{(1)} \\
&\qquad\qquad+ a^3 h_\tau{}^\mu \delta_\xi h_{\mu\tau} \Big](s,w,\theta) \total s \\
&= \int_{-\infty}^\tau \bigg[ \left( a \partial_\mu \xi^w + \partial_s \xi_\mu + \partial_\mu \xi_\tau - 2 \Gamma^\rho_{\tau\mu} \xi_\rho \right) \\
&\qquad\qquad\times \left( g^{\mu\nu} \partial_\nu w_{(1)} + a h_\tau{}^\mu \right) \bigg](s,w,\theta) \total s + \bigo{\kappa} \\
&= \int_{-\infty}^\tau \left[ \partial_s \xi^\mu \left( \partial_\mu w_{(1)} + a h_{\tau\mu} \right) + H \xi^\tau \left( \partial_s w_{(1)} + a h_{\tau\tau} \right) \right](s,w,\theta) \total s + \bigo{\kappa} \\
&= \int_{-\infty}^\tau \left[ \partial_s \left( \xi^\mu \partial_\mu w_{(1)} \right) + \frac{a}{2} \lie_\xi \left( a^2 h_{\mu\nu} \right) k^\mu k^\nu \right](s,w,\theta) \total s + \bigo{\kappa} \eqend{,}
\end{splitequation}
where we also used the relation~\eqref{eq:eom_1_w}. Similar expressions follow for the other second-order corrections, and the final result reads
\begin{equations}
\delta_\xi \tau_{(2)}(\tau,w,\theta) &= \xi^\mu \partial_\mu \tau_{(1)} + \frac{1}{2} \int_{-\infty}^\tau \left[ \lie_\xi \left( a^2 h_{\mu\nu} \right) u^\mu u^\nu \right]\left[s,w - \eta(\tau) + \eta(s),\theta\right] \total s \eqend{,} \\
\delta_\xi w_{(2)}(\tau,w,\theta) &= \xi^\mu \partial_\mu w_{(1)} + \frac{1}{2} \int_{-\infty}^\tau \left[ a \lie_\xi \left( a^2 h_{\mu\nu} \right) k^\mu k^\nu \right](s,w,\theta) \total s \eqend{,} \\
\delta_\xi \theta_{(2)}^a(\tau,w,\theta) &= \xi^\mu \partial_\mu \theta_{(1)}^a + \int_{-\infty}^\tau \left[ a \gamma^{ab} \lie_\xi \left( a^2 h_{\mu b} \right) k^\mu \right](s,w,\theta) \total s \eqend{.}
\end{equations}
Taking all together, the gauge transformation of the perturbed GLC coordinates yields
\begin{equations}
\delta_\xi \tilde{\tau} &= \kappa \xi^\tau + \kappa^2 \xi^\mu \partial_\mu \tau_{(1)} + \bigo{\kappa^3} = \kappa \xi^\mu \partial_\mu \tilde{\tau} + \bigo{\kappa^3} \eqend{,} \\
\delta_\xi \tilde{w} &= \kappa \xi^w + \kappa^2 \xi^\mu \partial_\mu w_{(1)} + \bigo{\kappa^3} = \kappa \xi^\mu \partial_\mu \tilde{w} + \bigo{\kappa^3} \eqend{,} \\
\delta_\xi \tilde{\theta}^a &= \kappa \xi^a + \kappa^2 \xi^\mu \partial_\mu \theta_{(1)}^a + \bigo{\kappa^3} = \kappa \xi^\mu \partial_\mu \tilde{\theta}^a + \bigo{\kappa^3} \eqend{,}
\end{equations}
showing that indeed the perturbed GLC coordinates transform as scalar fields under gauge transformations that preserve the background.

\subsection{Gauge freedom}
\label{sec:gauge_freedom}

The metric in the GLC coordinates~\eqref{eq:glc_metric} still has some remaining gauge freedom, even though it is parametrised by $n(n-1)/2$ independent functions~\cite{fanizza_etal_jacp_2013,fleury_nugier_fanizza_jcap_2016,fanizza_etal_jacp_2019}. This remaining gauge freedom consists of the relabeling of the lightcones by taking $w = w(w')$, or of the light rays by taking $\theta^a = \theta^a(w,\theta')$. Indeed, although the relabeling of the lightcones changes $\Upsilon$ and the relabeling of the light rays changes $U^a$ and $\gamma_{ab}$, these transformations keep the overall form of the GLC metric~\eqref{eq:glc_metric} unchanged. To completely fix this freedom, it is thus necessary to specify exactly how the coordinates $w$ and $\theta^a$ are defined by a given family of observers.

This is precisely the case for the GLC coordinates defined on the background FLRW spacetime~\eqref{eq:flrw_metric_glc}. The choice of the lightcone coordinate $w$ in Eq.~\eqref{eq:w_coord_flrw} and of the angular coordinates completely fix the remaining freedom in the GLC coordinates. The coordinate $w$ corresponds to the conformal reception time of photons by the observer sitting at $r = 0$, while the angular coordinates parametrise the unit sphere centered on that observer.

The remaining freedom of the GLC coordinates is also completely fixed at the perturbative level in our construction. Indeed, for a given metric perturbation $h_{\mu\nu}$ of compact support or fast fall-off, the perturbative corrections to the GLC coordinates found in Eqs.~\eqref{eq:X_1_integral} and~\eqref{eq:X_2_integral} are completely determined by choosing the perturbed GLC coordinates to coincide with the background ones in the absence of perturbations, and no relabeling of the lightcones or the light rays is possible anymore. Of course, the metric perturbation itself is not fixed, as it transforms under diffeomorphisms that leave the background invariant, according to Eq.~\eqref{eq:h_mu_nu_gauge_transformation}. As we saw in the previous section, these diffeomorphisms result in transformations of the perturbed GLC coordinates, which change as scalars.

We can take advantage of the gauge freedom of $h_{\mu\nu}$ to considerably simplify the form of the perturbed GLC coordinates, given in an arbitrary gauge in Eqs.~\eqref{eq:X_1_integral} and~\eqref{eq:X_2_integral}. A particularly interesting gauge choice is the one that eliminates all the non-local terms involving $h_{\mu\nu}$ appearing in those equations. This is achieved by requiring the background and perturbed GLC coordinates to coincide, \ie, taking GLC gauge for the full perturbed metric $\tilde{g}_{\mu\nu}$. Since the perturbed coordinates satisfy
\begin{equation}
\tilde{g}^{\mu\nu} \partial_\mu \tilde{\tau} \, \partial_\nu \tilde{\tau} = -1 \eqend{,} \qquad \tilde{g}^{\mu\nu} \partial_\mu \tilde{w} \, \partial_\nu \tilde{w} = 0 \eqend{,} \qquad \tilde{g}^{\mu\nu} \partial_\mu \tilde{w} \, \partial_\nu \tilde{\theta}^a = 0 \eqend{,}
\end{equation}
this choice amounts to imposing the following non-linear conditions:
\begin{equation}
\label{eq:gauge_cond_full}
\tilde{g}^{\mu\nu} \partial_\mu \tau \, \partial_\nu \tau = \tilde{g}^{\tau\tau} = - 1 \eqend{,} \qquad \tilde{g}^{\mu\nu} \partial_\mu w \, \partial_\nu w = \tilde{g}^{ww} = 0 \eqend{,} \qquad \tilde{g}^{\mu\nu} \partial_\mu w \, \partial_\nu \theta^a = \tilde{g}^{wa} = 0 \eqend{.}
\end{equation}
In this fully non-linear GLC gauge, the coordinate $\tau$ corresponds to cosmic time, while the angular coordinates $\theta^a$ are identified with the angular directions as seen by the observer sitting at $r = 0$. At linear order, the conditions~\eqref{eq:gauge_cond_full} assume the simpler form
\begin{equation}
\label{eq:gauge_cond_linear}
h_{\tau\tau} = h_{\tau a} = h_{ww} + 2 a h_{\tau w} = 0 \eqend{,}
\end{equation}
yielding $\tau_{(1)} = w_{(1)} = \theta^a_{(1)} = 0$ in the general expression~\eqref{eq:X_1_integral} as expected.

Interestingly, the linearised gauge-fixing conditions~\eqref{eq:gauge_cond_linear} are closely related to the ``observational synchronous gauge'' proposed in Ref.~\cite{fanizza_etal_jcap_2021}. The main difference between the perturbative corrections to the GLC coordinates~\eqref{eq:X_1_integral} that we define and the ones of Ref.~\cite{fanizza_etal_jcap_2021} is the choice of integration limits. In our case, as explained above we have fixed the remaining freedom in the choice of GLC coordinates by requiring the perturbed GLC coordinates to coincide with the background ones in the absence of perturbations, in particular when going to spatial or past infinity. In our opinion, this is the right choice for studying the effects of fluctuations throughout the universe, for example when computing quantum corrections, irrespectively of where the observer is located. On the other hand, the choice of Ref.~\cite{fanizza_etal_jcap_2021} corresponds to a more observer-centric view, and is better suited if one considers the effects fluctuations have on signals that reach a fixed observer. There is then still the freedom of fixing the GLC frame at the observer, corresponding to free integration constants, which can of course be chosen in such a way as to make their choice equal to ours.

\section{Gauge-invariant metric}                                                                                                                %
\label{sec:gauge_invariant_glc_metric}                                                                                                              %

The perturbative expansion of the invariant metric $\mathcal{G}_{\mu\nu}$ defined in Eq.~\eqref{eq:inv_metric} was given up to second order in terms of the metric perturbation $g^{(1)}_{\mu\nu} = a^2 h_{\mu\nu}$ in Eqs.~\eqref{eq:G_1} and~\eqref{eq:G_2}. In this section, we show that the invariant metric has the form of a metric in GLC coordinates, \ie, that
\begin{equation}
\label{eq:glc_metric_perturbed}
\mathcal{G}_{\mu\nu} = \begin{pmatrix} 0 & - \tilde{\Upsilon} & 0 \\ - \tilde{\Upsilon} & \tilde{\Upsilon}^2 + \tilde{\gamma}_{ab} \tilde{\mathsf{U}}^a \tilde{\mathsf{U}}^b & - \tilde{\gamma}_{bc} \tilde{\mathsf{U}}^c \\ 0 & - \tilde{\gamma}_{ac} \tilde{\mathsf{U}}^c & \tilde{\gamma}_{ab} \end{pmatrix} \eqend{,}
\end{equation}
where the perturbed GLC scalars $\tilde{\Upsilon}$ and $\tilde{\mathsf{U}}^a$ and the perturbed celestial sphere metric $\tilde{\gamma}_{ab}$ are all gauge-invariant objects.

We start by finding out how to express $\tilde{\Upsilon}$ and $\tilde{\mathsf{U}}^a$ in terms of the metric perturbation $h_{\mu\nu}$ in an arbitrary gauge. Let us first define the following scalars in the background coordinates:
\begin{equations}[eq:Y_u]
\tilde{Y}(x) &\equiv \left[ \left( \tilde{g}^{\mu\nu} \tilde{k}_\mu \tilde{u}_\nu \right)(x) \right]^{-1} \eqend{,} \label{eq:Y} \\
\tilde{U}^a(x) &\equiv \left( \tilde{Y} \tilde{g}^{\mu\nu} \tilde{u}_\mu \partial_\nu \tilde{\theta}^a \right)(x) \label{eq:U} \eqend{.}
\end{equations}
From the definition of the GLC scalars~\eqref{eq:Upsilon} and~\eqref{eq:U_a}, we see that $\tilde{Y}$ and $\tilde{U}^a$ correspond to the scalars $\tilde{\Upsilon}$ and $\tilde{\mathsf{U}}^a$, but written in the background coordinates. For later use, we expand them up to second order in perturbation theory. The expansion of $\tilde{Y}$ gives
\begin{equation}
\label{eq:Y_expansion}
\tilde{Y}(x) = a + \kappa Y^{(1)}(x) + \kappa^2 Y^{(2)}(x) + \bigo{\kappa^3}
\end{equation}
with
\begin{equations}
Y^{(1)} &= - a^2 \left[ a^{-1} \partial_\tau \tau_{(1)} + \left( \partial_\tau + a^{-1} \partial_w \right) w_{(1)} + a^2 h^{\tau w} \right] \eqend{,} \label{eq:Y_1} \\
\begin{split}
Y^{(2)} &= - a^2 \Big[ a^{-1} \partial_\tau \tau_{(2)} + \left( \partial_\tau + a^{-1} \partial_w \right) w_{(2)} - g^{\mu\nu} \partial_\mu \tau_{(1)} \partial_\nu w_{(1)} \\
&\qquad\quad+ a^2 h^{w\mu} \partial_\mu \tau_{(1)} + a^2 h^{\tau\mu} \partial_\mu w_{(1)} - a^4 h^{\tau\sigma} h_\sigma^{\phantom{\sigma} w} \Big] \\
&\quad+ a^3 \left[ a^{-1} \partial_\tau \tau_{(1)} + \left( \partial_\tau + a^{-1} \partial_w \right) w_{(1)} + a^2 h^{\tau w} \right]^2 \eqend{,}
\end{split} \label{eq:Y_2}
\end{equations}
where the first-order coordinate corrections $\tau_{(1)}$, $w_{(1)}$ and $\theta^a_{(1)}$ were given in Eq.~\eqref{eq:X_1_integral}, and the second-order ones in Eq.~\eqref{eq:X_2_integral}. The expansion of $\tilde{U}^a$ yields
\begin{equation}
\label{eq:u_a}
\tilde{U}^a(x) = \kappa U_{(1)}^a(x) + \kappa^2 U_{(2)}^a(x) + \bigo{\kappa^3}
\end{equation}
with
\begin{equations}
U_{(1)}^a &= - a \left[ \gamma^{ab} \partial_b \tau_{(1)} - \left( \partial_\tau + a^{-1} \partial_w \right) \theta_{(1)}^a - a^2 h^{\tau a} \right] \eqend{,} \label{eq:u_a_1} \\
\begin{split}
U_{(2)}^a &= - a \Big[ \gamma^{ab} \partial_b \tau_{(2)} - \left( \partial_\tau + a^{-1} \partial_w \right) \theta_{(2)}^a + g^{\mu\nu} \partial_\mu \tau_{(1)} \partial_\nu \theta_{(1)}^a - a^2 h^{a\mu} \partial_\mu \tau_{(1)} - a^2 h^{\tau\mu} \partial_\mu \theta_{(1)}^a \\
&\qquad\quad+ a^4 h^{\tau\sigma} h_\sigma{}^a \Big] - Y^{(1)} \left[ \gamma^{ab} \partial_b \tau_{(1)} - \left( \partial_\tau + a^{-1} \partial_w \right) \theta_{(1)}^a - a^2 h^{\tau a} \right] \eqend{.}
\end{split} \label{eq:u_a_2} \raisetag{1.6em}
\end{equations}

To find the invariant scalars appearing in the invariant metric~\eqref{eq:glc_metric_perturbed}, we have to transform $\tilde{Y}$ and $\tilde{U}^a$ to the GLC coordinates $\tilde{X}^{(\mu)} = (\tilde{\tau},\tilde{w},\tilde{\theta}^a)$. Hence, we use Eq.~\eqref{eq:gauge_invariant_tensor} and write
\begin{equations}
\tilde{\Upsilon}(\tilde{X}) &= \tilde{Y}[x(\tilde{X})] \eqend{,} \\
\tilde{\mathsf{U}}^a(\tilde{X}) &= \tilde{U}^a[x(\tilde{X})] \eqend{.}
\end{equations}
Their perturbative expansion is performed as in Eqs.~\eqref{eq:expansion_S}--\eqref{eq:scalar_observable_expansion}. The expansion of $\tilde{\Upsilon}$ yields
\begin{equation}\label{eq:Upsilon_expansion}
\tilde{\Upsilon}(\tilde{X}) = a + \kappa \Upsilon^{(1)}(\tilde{X}) + \kappa^2 \Upsilon^{(2)}(\tilde{X}) + \bigo{\kappa^3}
\end{equation}
with
\begin{equations}
\Upsilon^{(1)} &= Y^{(1)} - H a \tau_{(1)} \eqend{,} \label{eq:Upsilon_1} \\
\Upsilon^{(2)} &= Y^{(2)} - X^{(\mu)}_{(1)} \partial_\mu Y^{(1)} - H a \left( \tau_{(2)} - X^{(\mu)}_{(1)} \partial_\mu \tau_{(1)} \right) + \frac{1}{2} (1-\epsilon) H^2 a \tau_{(1)}^2 \eqend{,} \label{eq:Upsilon_2}
\end{equations}
and where $X_{(1)}^{(\mu)} = \left( \tau_{(1)}, w_{(1)}, \theta_{(1)}^a \right)$. The expansion of $\tilde{\mathsf{U}}^a$ reads
\begin{equation}
\label{eq:U_a_expansion}
\tilde{\mathsf{U}}^a(\tilde{X}) = \kappa \mathsf{U}^a_{(1)}(\tilde{X}) + \kappa^2 \mathsf{U}^a_{(2)}(\tilde{X}) + \bigo{\kappa^3}
\end{equation}
with
\begin{equations}
\mathsf{U}^a_{(1)} &= U^a_{(1)} \eqend{,} \label{eq:U_a_1} \\
\mathsf{U}^a_{(2)} &= U^a_{(2)} - X^{(\mu)}_{(1)} \partial_\mu U^a_{(1)} \eqend{.} \label{eq:U_a_2}
\end{equations}
We note that the scalars $\tilde{\Upsilon}$ and $\tilde{\mathsf{U}}^a$ can be easily related to the components of the inverse invariant metric $\mathcal{G}^{\mu\nu}$ defined in Eq.~\eqref{eq:inverse_inv_metric}, since
\begin{equation}
\frac{1}{\tilde{\Upsilon}(\tilde{X})} = \left( \tilde{g}^{\mu\nu} \tilde{k}_\mu \tilde{u}_\nu \right)[x(\tilde{X})] = - \frac{\partial \tilde{\tau}}{\partial x^\mu} \frac{\partial \tilde{w}}{\partial x^\nu} \tilde{g}^{\mu\nu}[x(\tilde{X})] = - \mathcal{G}^{\tau w}(\tilde{X})
\end{equation}
and
\begin{equation}
\frac{\tilde{\mathsf{U}}^a(\tilde{X})}{\tilde{\Upsilon}(\tilde{X})} = \left( \tilde{g}^{\mu\nu} \tilde{u}_\mu \partial_\nu \tilde{\theta}^a \right)[x(\tilde{X})] = - \frac{\partial \tilde{\tau}}{\partial x^\mu} \frac{\partial \tilde{\theta}^a}{\partial x^\nu} \tilde{g}^{\mu\nu}[x(\tilde{X})] = - \mathcal{G}^{\tau a}(\tilde{X}) \eqend{.}
\end{equation}
This shows again the advantage of our geometrical construction of invariant observables, using a field-dependent diffeomorphism: relations between tensors that are fulfilled in the perturbed geometry are also fulfilled for the invariant observables corresponding to these tensors.

The expression for $\tilde{\gamma}_{ab}$ is obtained straightforwardly from the transformation of the perturbed metric $\tilde{g}_{\mu\nu}$ from the background to the perturbed GLC coordinates, see Eq.~\eqref{eq:celestial_sphere_metric}. We simply have
\begin{equation}
\tilde{\gamma}_{ab}(\tilde{X}) = \frac{\partial x^\mu}{\partial \tilde{\theta}^a} \frac{\partial x^\nu}{\partial \tilde{\theta}^b} \tilde{g}_{\mu\nu}[x(\tilde{X})] = \mathcal{G}_{ab}(\tilde{X})\eqend{,}
\end{equation}
where we have used Eq.~\eqref{eq:inv_metric} in the second equality. For later use, we also write down the perturbative series for $\tilde{\gamma}_{ab}$, which reads
\begin{equation}
\tilde{\gamma}_{ab} = \gamma_{ab} + \kappa \gamma_{ab}^{(1)} + \kappa^2 \gamma_{ab}^{(2)} + \bigo{\kappa^3} \eqend{,}
\end{equation}
and whose components may be evaluated explicitly using the expressions~\eqref{eq:G_1} and~\eqref{eq:G_2} for the invariant metric.

\subsection{First order}

To compute the first-order corrections to the invariant metric $\mathcal{G}_{\mu\nu}$, we start from Eq.~\eqref{eq:G_1}. Then, after using the differential equations~\eqref{eq:eom_1} for the first-order corrections to the perturbed GLC coordinates to make some simplifications, we obtain
\begin{equations}[eq:G_mu_nu_1]
\mathcal{G}_{\tau\tau}^{(1)} &= 0 \eqend{,} \\
\mathcal{G}_{\tau w}^{(1)} &= a^2 h_{\tau w} + \frac{1}{2} a^3 h_{\tau\tau} + a \partial_\tau \tau_{(1)} + a \partial_w w_{(1)} + H a \tau_{(1)} \eqend{,} \\
\mathcal{G}_{\tau a}^{(1)} &= 0 \eqend{,} \\
\mathcal{G}_{ww}^{(1)} &= - 2 a \, \mathcal{G}_{\tau w}^{(1)} \eqend{,} \\
\mathcal{G}_{wa}^{(1)} &= a^2 h_{wa} - \gamma_{ab} \partial_w \theta_{(1)}^b + a \partial_a \tau_{(1)} - a^2 \partial_a w_{(1)} \eqend{,} \\
\mathcal{G}_{ab}^{(1)} &= a^2 h_{ab} - 2 \gamma_{ab} \left[ H - \frac{1}{a(w - \eta)} \right] \tau_{(1)} - 2 \gamma_{ab} \frac{w_{(1)}}{w - \eta} - \theta_{(1)}^c \partial_c \gamma_{ab} - 2 \gamma_{c(a} \partial_{b)} \theta_{(1)}^c \eqend{.}
\end{equations}

Next, we will find explicit formulas for the first-order corrections to $\tilde{\Upsilon}$ and $\tilde{\mathsf{U}}^a$. From Eqs.~\eqref{eq:Y_1} and~\eqref{eq:Upsilon_1}, we find that the correction to $\tilde{\Upsilon}$ has the form
\begin{equation}
\label{eq:Upsilon_1_final}
\Upsilon^{(1)} = - \left( a^2 h_{\tau w} + \frac{1}{2} a^3 h_{\tau\tau} + a \partial_\tau \tau_{(1)} + a \partial_w w_{(1)} + H a \tau_{(1)} \right) \eqend{,}
\end{equation}
after using the differential equations~\eqref{eq:eom_1}. The first-order correction to $\tilde{\mathsf{U}}^a$ is
\begin{equation}
\label{eq:U_a_1_final}
\mathsf{U}^a_{(1)} = - \gamma^{ab} \left( a^2 h_{wb} - \gamma_{bc} \partial_w \theta_{(1)}^c + a \partial_b \tau_{(1)} - a^2 \partial_b w_{(1)} \right) \eqend{,}
\end{equation}
which is obtained by combining Eqs.~\eqref{eq:u_a_1} and~\eqref{eq:U_a_1}, and simplified with the help of Eq.~\eqref{eq:eom_1}. Hence, comparing Eq.~\eqref{eq:G_mu_nu_1} with the results above, we conclude that
\begin{equation}
\label{eq:G_mu_nu_1_glc}
\mathcal{G}_{\tau w}^{(1)} = - \Upsilon^{(1)} \eqend{,} \quad \mathcal{G}_{ww}^{(1)} = 2 a \Upsilon^{(1)} \eqend{,} \quad \mathcal{G}_{wa}^{(1)} = - \gamma_{ab} \mathsf{U}_{(1)}^b \eqend{,} \quad \mathcal{G}_{ab}^{(1)} = \gamma^{(1)}_{ab} \eqend{.}
\end{equation}
These corrections can also be written in the more compact form
\begin{equation}
\label{eq:gauge_inv_metric_perturbation_glc}
\mathcal{G}_{\mu\nu}^{(1)} = 2 \left( u_{(\mu} k_{\nu)} + a k_\mu k_\nu \right) \Upsilon^{(1)} - 2 k_{(\mu} \gamma_{\nu)\rho} \mathsf{U}^{\rho}_{(1)} + \gamma_{\mu\nu}^{(1)} \eqend{,}
\end{equation}
it being understood that $\gamma_{\mu\nu}^{(1)}$ has only angular components:
\begin{equation}
u^\mu \gamma_{\mu\nu}^{(1)} = k^\mu \gamma_{\mu\nu}^{(1)} = 0 \eqend{.}
\end{equation}

It is also interesting to express the metric perturbation $h_{\mu\nu}$ in terms of the gauge-independent corrections $\Upsilon^{(1)}$ and $\mathsf{U}_{(1)}^a$ we have just computed. The gauge dependence of $h_{\mu\nu}$, however, means that we will also need the corrections to the GLC coordinates. Thus, by combining Eq.~\eqref{eq:eom_1} and Eqs.~\eqref{eq:G_mu_nu_1}--\eqref{eq:U_a_1_final}, we can express the components of $h_{\mu\nu}$ as
\begin{equations}[eq:h_mu_nu_first_order_glc]
h_{\tau\tau} &= - 2 a^{-1} \partial_\tau w_{(1)} \eqend{,} \\
h_{\tau w} &= - a^{-2} \Upsilon^{(1)} - H a^{-1} \tau_{(1)} - a^{-1} \partial_\tau \tau_{(1)} + \left( \partial_\tau - a^{-1} \partial_w \right) w_{(1)} \eqend{,} \\
h_{\tau a} &= - a^{-1} \partial_a w_{(1)} + a^{-2} \gamma_{ab} \partial_\tau \theta_{(1)}^b \eqend{,} \\
h_{ww} &= 2 a^{-1} \Upsilon^{(1)} + 2 H \tau_{(1)} - 2 a^{-1} \partial_w \tau_{(1)} + 2 \partial_w w_{(1)} \eqend{,} \\
h_{wa} &= - a^{-2} \gamma_{ab} \mathsf{U}_{(1)}^b + a^{-2} \gamma_{ab} \partial_w \theta_{(1)}^b - a^{-1} \partial_a \tau_{(1)} + \partial_a w_{(1)} \eqend{,} \\
h_{ab} &= a^{-2} \gamma_{ab}^{(1)} + 2 a^{-2} \gamma_{ab} \left[ \left( H - \frac{1}{a (w-\eta)} \right) \tau_{(1)} + \frac{w_{(1)}}{w-\eta} \right] + a^{-2} \theta_{(1)}^c \partial_c \gamma_{ab} + 2 a^{-2} \gamma_{c(a} \partial_{b)} \theta_{(1)}^c \eqend{.}
\end{equations}
In this way, it is easy to read off the gauge-invariant and gauge-dependent parts of the metric perturbation at first order: the gauge-dependent parts are the ones that depend on the first-order coordinate corrections $\tau_{(1)}$, $w_{(1)}$ and $\theta^a_{(1)}$, while the gauge-invariant parts involve $\Upsilon^{(1)}$, $\mathsf{U}_{(1)}^a$ and $\gamma_{ab}^{(1)}$. We note that we can write Eq.~\eqref{eq:h_mu_nu_first_order_glc} as
\begin{equation}
h_{\mu\nu} = a^{-2} \mathcal{G}_{\mu\nu}^{(1)} + 2 a^{-2} g_{\rho(\mu} \nabla_{\nu)} X^{(\rho)}_{(1)} \eqend{,}
\end{equation}
which together with the first-order transformation~\eqref{eq:X_1_gauge_transformation} of the perturbed coordinates exactly gives the transformation~\eqref{eq:h_mu_nu_gauge_transformation} for the metric perturbation (at first order).

\subsection{Second order}

At second order, the explicit forms of $\mathcal{G}_{\mu\nu}^{(2)}$ in terms of $g^{(1)}_{\mu\nu} = a^2 h_{\mu\nu}$ are considerably more involved than its first-order counterpart and not particularly illuminating, see Eq.~\eqref{eq:G_2}. Here we are interested in checking that $\mathcal{G}_{\mu\nu}$ has the form~\eqref{eq:glc_metric_perturbed} also at second order, rather than the explicit formulas involving $h_{\mu\nu}$. Therefore, we shall restrict ourselves to the task of verifying the vanishing components of the invariant metric and its expression in terms of the scalars $\tilde{\Upsilon}$ and $\tilde{\mathsf{U}}^a$ at second order in perturbation theory.

We start by showing that the corrections to $\mathcal{G}_{\tau\tau}$ and $\mathcal{G}_{\tau a}$ also vanish at second order. From Eq.~\eqref{eq:G_2} we have that the second-order correction to the time-time component of the invariant metric is given by
\begin{splitequation}
\mathcal{G}_{\tau\tau}^{(2)} &= - a^2 \left( \tau_{(1)} \partial_\tau + w_{(1)} \partial_w + \theta^a_{(1)} \partial_a + 2 H \tau_{(1)} + 2 \partial_\tau \tau_{(1)} \right) h_{\tau\tau} - 2 \left( a^2 h_{\tau w} + H a \tau_{(1)} \right) \partial_\tau w_{(1)} \\
&\quad- 2 a^2 h_{\tau a} \partial_\tau \theta^a_{(1)} - 2 a \partial_\tau \tau_{(1)} \partial_\tau w_{(1)} + a^2 \left( \partial_\tau w_{(1)} \right)^2 + \gamma_{ab} \partial_\tau \theta^a_{(1)} \partial_\tau \theta^b_{(1)} \\
&\quad+ 2 a \partial_\tau \left( w_{(2)} - \tau_{(1)} \partial_\tau w_{(1)} - w_{(1)} \partial_w w_{(1)} - \theta^a_{(1)} \partial_a w_{(1)} \right) \\
&= 2 a \partial_\tau w_{(2)} - a^2 \left( g^{\mu\nu} \partial_\mu w_{(1)} \partial_\nu w_{(1)} + 2 a h_\tau{}^\nu \partial_\nu w_{(1)} + a^2 h_\tau{}^\nu h_{\nu\tau} \right) = 0 \eqend{,} \raisetag{1.6em}
\end{splitequation}
where we have used Eq.~\eqref{eq:eom_1} in the first equality and Eq.~\eqref{eq:eom_2} to obtain the final result. The other vanishing second-order correction to the invariant metric is
\begin{splitequation}
\mathcal{G}^{(2)}_{\tau a} &= - a^2 \left( \tau_{(1)} \partial_\tau + w^{(1)} \partial_w + \theta^b_{(1)} \partial_b + 2 H \tau_{(1)} \right) h_{\tau a} - a \partial_\tau \tau_{(1)} \partial_a w_{(1)} - a \partial_\tau w_{(1)} \partial_a \tau_{(1)} \\
&\quad+ a^2 \partial_\tau w_{(1)} \partial_a w_{(1)} + \gamma_{bc} \partial_\tau \theta^b_{(1)} \partial_a \theta^c_{(1)} - a^2 \left( \partial_\tau \tau_{(1)} h_{\tau a} + \partial_\tau w_{(1)} h_{w a} + \partial_\tau \theta^b_{(1)} h_{ba} \right) \\
&\quad+ \left( \tau_{(1)} \partial_\tau + w_{(1)} \partial_w \right) \gamma_{ab} \partial_\tau \theta^b_{(1)} - a^2 \left( \partial_a \tau_{(1)} h_{\tau\tau} + \partial_a w_{(1)} h_{\tau w} + \partial_a \theta^b_{(1)} h_{\tau b} \right)\\
&\quad- H a \tau_{(1)} \partial_a w_{(1)} - \gamma_{ab} \partial_\tau \theta^b_{(2)} + \gamma_{ab} \partial_\tau \left( \tau_{(1)} \partial_\tau + w^{(1)} \partial_w + \theta^c_{(1)} \partial_c \right) \theta_{(1)}^b \\
&\quad+ a \partial_a w_{(2)} - a \partial_a \left( \tau_{(1)} \partial_\tau + w^{(1)} \partial_w + \theta^b_{(1)} \partial_b \right) w_{(1)} \\
&= - \gamma_{ab} \partial_\tau \theta^b_{(2)} + a \partial_a w_{(2)} + a \gamma_{ab} g^{\mu\nu} \partial_\mu w_{(1)} \partial_\nu \theta_{(1)}^b + a^2 \gamma_{ab} h_\tau{}^\nu \partial_\nu \theta_{(1)}^b \\
&\quad- a^3 h_a{}^\nu \partial_\nu w_{(1)} - a^4 h_\tau{}^\nu h_{\nu a} = 0 \eqend{,}
\end{splitequation}
where again we have made use of Eqs.~\eqref{eq:eom_1} and~\eqref{eq:eom_2}.

Our next step is to verify that $\Upsilon^{(2)}$ in Eq.~\eqref{eq:Upsilon_2} corresponds to (minus) the second-order correction to $\mathcal{G}_{\tau w}$. We first rewrite the terms involving the time derivative of the scale factor, using that $g_{\tau w} = - a$, to obtain
\begin{equation}
\label{eq:Upsilon_2_partial}
\Upsilon^{(2)} = Y^{(2)} - X^{(\rho)}_{(1)} \partial_\rho Y^{(1)} + \left[ X^{(\rho)}_{(2)} - X^{(\sigma)}_{(1)} \partial_\sigma X^{(\rho)}_{(1)} \right] \partial_\rho g_{\tau w} - \frac{1}{2}X^{(\rho)}_{(1)} X^{(\sigma)}_{(1)} \partial_\rho \partial_\sigma g_{\tau w} \eqend{.}
\end{equation}
Using Eqs.~\eqref{eq:Y_1} and~\eqref{eq:eom_1}, we then find
\begin{equation}
\label{eq:Y_1_partial}
Y^{(1)} = - a^2 h_{\tau w} + \partial_\tau X^{(\rho)}_{(1)} g_{\rho w} + \partial_w X^{(\rho)}_{(1)} g_{\tau \rho} \eqend{.}
\end{equation}
For $Y^{(2)}$ we use Eq.~\eqref{eq:Y_2} and obtain
\begin{splitequation}
\label{eq:Y_2_partial}
Y^{(2)} &= a^4 \left( \partial_\tau X^{(\rho)}_{(2)} g_{\rho w} + \partial_w X^{(\rho)}_{(2)} g_{\tau\rho} \right) - a \partial_\tau \tau_{(1)} \partial_w w_{(1)} + a \partial_\tau \theta^a_{(1)} \partial_a \left( \tau_{(1)} - a w_{(1)} \right) \\
&\quad- a^2 h_{\tau w} \partial_\tau \tau_{(1)} + \frac{a^2}{2} \left( h_{ww} + 4 a h_{\tau w} \right) \partial_\tau w_{(1)} - a^2 h_{\tau w} \partial_w w_{(1)} + a^3 h_{wa} \partial_\tau \theta^a_{(1)} \\
&\quad- 2 a \left( \partial_\tau X^{(\rho)}_{(1)} g_{\rho w} - \partial_w X^{(\rho)}_{(1)} g_{\tau\rho} \right) h_{\tau w} + a^{-1} \left( \partial_\tau X^{(\rho)}_{(1)} g_{\rho w} - \partial_w X^{(\rho)}_{(1)} g_{\tau\rho} \right)^2 \eqend{,}
\end{splitequation}
again with the aid of Eqs.~\eqref{eq:eom_1} and~\eqref{eq:eom_2}. Using Eqs.~\eqref{eq:Y_1_partial} and~\eqref{eq:Y_2_partial} in the first two terms in Eq.~\eqref{eq:Upsilon_2_partial}, we can cast those terms into the form
\begin{splitequation}
\label{eq:first_two_terms}
Y^{(2)} - X^{(\rho)}_{(1)}\partial_\rho Y^{(1)} &= \partial_\tau X^{(\rho)}_{(1)} \left( a^2 h_{\rho w} - X^{(\sigma)}_{(1)} \partial_\sigma g_{\rho w} \right) + \partial_w X^{(\rho)}_{(1)} \left( a^2 h_{\tau\rho} - X^{(\sigma)}_{(1)} \partial_\sigma g_{\tau\rho} \right) \\
&\quad+ X^{(\rho)}_{(1)} \partial_\rho \left( a^2 h_{\tau w} \right) - \partial_\tau X^{(\rho)}_{(1)} \partial_w X^{(\sigma)}_{(1)} g_{\rho\sigma} \\
&\quad+ \partial_\tau \left( X^{(\rho)}_{(2)} - X^{(\sigma)}_{(1)} \partial_\sigma X^{(\rho)}_{(1)} \right) g_{\rho w} + \partial_w \left( X^{(\rho)}_{(2)} - X^{(\sigma)}_{(1)} \partial_\sigma X^{(\rho)}_{(1)} \right) g_{\tau\rho} \eqend{,}
\end{splitequation}
using again Eqs.~\eqref{eq:eom_1} and~\eqref{eq:eom_2}. Finally, combining Eqs.~\eqref{eq:Upsilon_2_partial} and~\eqref{eq:first_two_terms}, and comparing the result with Eq.~\eqref{eq:G_2} we conclude that
\begin{equation}
\Upsilon^{(2)} = - \mathcal{G}_{\tau w}^{(2)} \eqend{.}
\end{equation}

We now turn to the scalars $\tilde{\mathsf{U}}^a$, which are related to $\mathcal{G}_{wa}$ as $\tilde{\gamma}_{ab} \tilde{\mathsf{U}}^b = - \mathcal{G}_{wa}$, see Eq.~\eqref{eq:glc_metric_perturbed}. To check their relation at second order in perturbation theory, we first notice that the second-order correction to $\tilde{\gamma}_{ab} \tilde{\mathsf{U}}^b$ is
\begin{equation}
\label{eq:gamma_ab_U_b_2}
\left( \tilde{\gamma}_{ab} \tilde{\mathsf{U}}^b \right)^{(2)} = \gamma_{ab} \mathsf{U}^b_{(2)} + \gamma_{ab}^{(1)} \mathsf{U}^b_{(1)} \eqend{.}
\end{equation}
The term $\mathsf{U}^a_{(2)}$ was given in Eq.~\eqref{eq:U_a_2} in terms of $U^a_{(1)}$ and $U^a_{(2)}$. It is convenient to first use Eqs.~\eqref{eq:eom_1}, \eqref{eq:eom_2} and~\eqref{eq:Y_1_partial} in the expression~\eqref{eq:u_a_2} for $U^a_{(2)}$ to obtain
\begin{splitequation}
\label{eq:gamma_ab_u_b_2}
\gamma_{ab} U^b_{(2)} &= \partial_a X^{(\mu)}_{(2)} g_{\mu w} + \partial_w X^{(\mu)}_{(2)} g_{\mu a} + a^2 \gamma_{ab} \left( \partial^\mu w_{(1)} - a^{-1} \partial^\mu \tau_{(1)} - h_w{}^\mu \right) \partial_\mu \theta^b_{(1)} \\
&\quad- a^4 h_a{}^\mu \partial_\mu w_{(1)} + a^3 h_a{}^\mu \partial_\mu \tau_{(1)} + a^4 h_w{}^\mu h_{\mu a} \\
&\quad+ a^{-1} \left( a^2 h_{\tau w} - \partial_\tau X^{(\mu)}_{(1)} g_{\mu w} - \partial_w X^{(\mu)}_{(1)} g_{\mu \tau} \right) \left( a^2 h_{wa} - \partial_a X^{(\nu)}_{(1)} g_{\nu w} - \partial_w X^{(\nu)}_{(1)} g_{\nu a} \right) \\
&= \partial_a X^{(\mu)}_{(2)} g_{\mu w} + \partial_w X^{(\mu)}_{(2)} g_{\mu a} - \partial_w X^{(\mu)}_{(1)} \partial_a X^{(\nu)}_{(1)} g_{\mu\nu} + a^2 h_{w\mu} \partial_a X^{(\mu)}_{(1)} - a^2 \partial_w w_{(1)} \partial_a w_{(1)} \\
&\quad+ \partial_w \theta^b_{(1)} \partial_a \theta^c_{(1)} \gamma_{bc} - a \partial_w w_{(1)} \partial_w \theta^b_{(1)} \gamma_{ab} - \left( \partial_w \tau_{(1)} + a^2 h_{\tau w} + \frac{a}{2} h_{ww} \right) \partial_a \tau_{(1)} \\
&\quad+ a^2 \left( \partial^c w_{(1)} - a^{-1} \partial^c \tau_{(1)} - h_w{}^c \right) \left( \partial_c \theta^b_{(1)} \gamma_{ab} - a^2 h_{ca} \right) - a^2 h_{wb} \partial_a \theta^b_{(1)} + a^2 h_{wa} \partial_w w_{(1)} \eqend{.}
\end{splitequation}
The term involving $U^a_{(1)}$ in Eq.~\eqref{eq:U_a_2} can be written as
\begin{splitequation}
\gamma_{ab} X^{(\mu)}_{(1)} \partial_\mu U^a_{(1)} &= - X^{(\mu)}_{(1)} \partial_\mu \left( a^2 h_{wa} \right) + X^{(\mu)}_{(1)} \partial_\mu \partial_w X^{(\nu)}_{(1)} g_{\nu a} + X^{(\mu)}_{(1)} \partial_\mu \partial_a X^{(\nu)}_{(1)} g_{\nu w} \\
&\quad+ \left( a^2 h_{wc} - \partial_c X^{(\nu)}_{(1)} g_{\nu w} - \partial_w X^{(\nu)}_{(1)} g_{\nu c} \right) \gamma^{cb} X^{(\mu)}_{(1)} \partial_\mu \gamma_{ab} \eqend{,}
\end{splitequation}
and for the remaining terms in Eq.~\eqref{eq:gamma_ab_U_b_2}, we see from Eq.~\eqref{eq:U_a_1_final} that $\mathsf{U}^b_{(1)}$ can be written as
\begin{equation}
\label{eq:U_a_1_closed form}
\mathsf{U}^a_{(1)} = - \gamma^{ab} \left( a^2 h_{wb} - \partial_w X^{(\mu)}_{(1)} g_{\mu b} - \partial_b X^{(\mu)}_{(1)} g_{w\mu} \right) \eqend{.}
\end{equation}
Moreover, the first-order correction to the celestial sphere metric $\tilde{\gamma}_{ab}$ was given in Eq.~\eqref{eq:G_mu_nu_1} and reads
\begin{equation}
\label{eq:gamma_ab_1}
\gamma^{(1)}_{ab} = a^2 h_{ab} - X^{(\mu)}_{(1)} \partial_\mu \gamma_{ab} - \partial_a X^{(\mu)}_{(1)} g_{\mu b} - \partial_b X^{(\mu)}_{(1)} g_{\mu a} \eqend{.}
\end{equation}
Taking all together, the substitution of Eqs.~\eqref{eq:gamma_ab_u_b_2}--\eqref{eq:gamma_ab_1} into Eq.~\eqref{eq:gamma_ab_U_b_2} results in
\begin{equation}
\left( \tilde{\gamma}_{ab} \tilde{\mathsf{U}}^b \right)^{(2)} = - \mathcal{G}_{wa}^{(2)} \eqend{.}
\end{equation}

Our final task is to check that the combination $\tilde{\Upsilon}^2 + \tilde{\gamma}_{ab} \tilde{\mathsf{U}}^a \tilde{\mathsf{U}}^b$ corresponds to $\mathcal{G}_{ww}$ at second order. From the perturbative expansion, we have
\begin{equation}
\left( \tilde{\Upsilon}^2 \right)^{(2)} = 2 a \Upsilon^{(2)} + \left( \Upsilon^{(1)} \right)^2 \eqend{,} \qquad \left( \tilde{\gamma}_{ab} \tilde{\mathsf{U}}^a \tilde{\mathsf{U}}^b \right)^{(2)} = \gamma_{ab} \mathsf{U}_{(1)}^a \mathsf{U}_{(1)}^b \eqend{.}
\end{equation}
We first treat the terms involving the perturbative expansion of $\tilde{\Upsilon}$. Using Eqs.~\eqref{eq:Upsilon_1} and \eqref{eq:Upsilon_2_partial} and the differential equations~\eqref{eq:eom_1} and~\eqref{eq:eom_2} for the corrections to the GLC coordinates, we find
\begin{splitequation}
\label{eq:Upsilon_part}
\left( \tilde{\Upsilon}^2 \right)^{(2)} &= 2 a \left[ Y^{(2)} - X^{(\rho)}_{(1)} \partial_\rho Y^{(1)} + \left[ X^{(\rho)}_{(2)} - X^{(\sigma)}_{(1)} \partial_\sigma X^{(\rho)}_{(1)} \right] \partial_\rho g_{\tau w} - \frac{1}{2} X^{(\rho)}_{(1)} X^{(\sigma)}_{(1)} \partial_\rho \partial_\sigma g_{\tau w} \right] \\
&\quad+ \left( Y^{(1)} - H a \tau_{(1)} \right)^2 \\
&= - 2 \partial_w \left[ X^{(\mu)}_{(2)} - X^{(\nu)}_{(1)} \partial_\nu X^{(\mu)}_{(1)} \right] g_{\mu w} + 2 \partial_w X^{(\mu)}_{(1)} X^{(\nu)}_{(1)} \partial_\nu g_{\mu w} - X^{(\mu)}_{(1)} \partial_\mu \left( a^2 h_{ww} \right) \\
&\quad- \left[ X^{(\rho)}_{(2)} - X^{(\sigma)}_{(1)} \partial_\sigma X^{(\rho)}_{(1)} \right] \partial_\rho g_{ww} + \frac{1}{2} X^{(\rho)}_{(1)} X^{(\sigma)}_{(1)} \partial_\rho \partial_\sigma g_{ww} - 2 \partial_w X^{(\mu)}_{(1)} \partial_\mu X^{(\nu)}_{(1)} g_{\nu w} \\
&\quad- a^2 g^{\mu\nu} \left( \partial_\mu \tau_{(1)} - a \partial_\mu w_{(1)} + 2 a h_{w\mu} \right) \left( \partial_\nu \tau_{(1)} - a \partial_\nu w_{(1)} \right) \\
&\quad+ 3 \left( \partial_w \tau_{(1)} - a \partial_w w_{(1)} + \frac{1}{2} a h_{ww} \right)^2 - a^4 h_w{}^\mu h_{\mu w} \eqend{.} \raisetag{1.8em}
\end{splitequation}
The term involving $\mathsf{U}^a_{(1)}$ is simpler and reads
\begin{splitequation}
\label{eq:U_part}
\gamma_{ab} \mathsf{U}^a_{(1)} \mathsf{U}^b_{(1)} &= a^4 h_w{}^a h_{aw} + \gamma_{ab} \partial_w \theta^a_{(1)} \partial_w \theta^b_{(1)} + a^2 \gamma^{ab} \left( \partial_a \tau_{(1)} - a \partial_a w_{(1)}\right) \left( \partial_b \tau_{(1)} - a \partial_b w_{(1)} \right) \\
&\quad- 2 a^2 h_{wa} \partial_w \theta^a_{(1)} - 2 a^2 h_w{}^a \partial_a X^{(\mu)}_{(1)} g_{\mu w} + 2 \partial_w \theta^a_{(1)} \partial_a X^{(\mu)}_{(1)} g_{\mu w} \eqend{.}
\end{splitequation}
Finally, combining Eqs.~\eqref{eq:Upsilon_part} and~\eqref{eq:U_part} and comparing the result with Eq.~\eqref{eq:G_2} we find
\begin{equation}
\left( \tilde{\Upsilon}^2 + \tilde{\gamma}_{ab} \tilde{\mathsf{U}}^a \tilde{\mathsf{U}}^b \right)^{(2)} = \mathcal{G}_{ww}^{(2)} \eqend{.}
\end{equation}

In conclusion, we have shown that the following equations hold at second order in perturbation theory:
\begin{equation}
\label{eq:G_mu_nu_2_glc}
\mathcal{G}_{\tau w}^{(2)} = - \Upsilon^{(2)} \eqend{,} \qquad \mathcal{G}_{ww}^{(2)} = 2 a \Upsilon^{(2)} + \left( \Upsilon^{(1)} \right)^2 + \gamma_{ab} \mathsf{U}_{(1)}^a \mathsf{U}_{(1)}^b \eqend{,} \qquad \mathcal{G}_{wa}^{(2)}  = - \mathsf{U}_a^{(2)} \eqend{.}
\end{equation}
That is, up to second order in perturbations we have explicitly verified that the invariant metric has the form of a metric in GLC coordinates~\eqref{eq:glc_metric_perturbed}.

\section{Linearised Einstein's equations}                                                                                                           %
\label{sec:linearised_einstein_eqs}                                                                                                                 %

In this section, we consider a single-field inflationary model, where inflation is driven by a minimally coupled scalar field $\tilde{\phi}$, the inflaton~\cite{mukhanov_feldman_brandenberger_pr_1992}. We still consider a FLRW background spacetime, such that the background inflaton field $\phi(\tau)$ only depends on time. We further make the standard assumption that it has everywhere timelike gradient, $\dot\phi < 0$. In addition to the metric pertubations~\eqref{eq:perturbed_metric_flrw}, also the inflaton has a perturbative expansion:
\begin{equation}
\label{eq:perturbed_inflaton}
\tilde{\phi} = \phi + \kappa \phi^{(1)} \eqend{.}
\end{equation}
For simplicity, in this section only, we also assume a four-dimensional spacetime.

\subsection{Equations for the linearised metric and inflaton}

Let us consider Einstein's equations for the perturbed spacetime,
\begin{equation}
\label{eq:einstein}
\tilde{G}_{\mu\nu} = \frac{\kappa^2}{2} \tilde{T}_{\mu\nu} \eqend{,}
\end{equation}
sourced by the energy-momentum tensor of the perturbed inflaton field:
\begin{equation}
\label{eq:inflaton_stresstensor}
\tilde{T}_{\mu\nu} = \partial_\mu \tilde{\phi} \partial_\nu \tilde{\phi} - \frac{1}{2} \tilde{g}_{\mu\nu} \left[ \tilde{g}^{\rho\sigma} \partial_\rho \tilde{\phi} \partial_\sigma \tilde{\phi} + V(\tilde{\phi}) \right] \eqend{,}
\end{equation}
where $V$ denotes the inflaton potential. We use Eqs.~\eqref{eq:perturbed_metric_flrw} and~\eqref{eq:perturbed_inflaton} to expand both the Einstein and the energy-momentum tensors up to first order in $\kappa$. Hence, we have
\begin{equations}
\tilde{G}_{\mu\nu} &= G_{\mu\nu} + \kappa G_{\mu\nu}^{(1)} + \bigo{\kappa^2} \eqend{,} \\
\tilde{T}_{\mu\nu} &= T_{\mu\nu} + \kappa T_{\mu\nu}^{(1)} + \bigo{\kappa^2} \eqend{.}
\end{equations}
For the Einstein tensor, we find
\begin{equations}
G_{\mu\nu} &= - (3-2 \epsilon) H^2 g_{\mu\nu} + 2 H^2 \epsilon u_\mu u_\nu \eqend{,} \\
\begin{split}
G_{\mu\nu}^{(1)} &= \nabla^\rho \nabla_{(\mu} \left( a^2 h_{\nu)\rho} \right) - \frac{1}{2} \nabla^2 \left( a^2 h_{\mu\nu} \right) - \frac{1}{2} g_{\mu\nu} \nabla^\rho \nabla^\sigma \left( a^2 h_{\rho\sigma} \right) - \frac{1}{2} \nabla_\mu \nabla_\nu \left( a^2 h \right) \\
&\quad+ \frac{1}{2} g_{\mu\nu} \nabla^2 \left( a^2 h \right) - 3 (2-\epsilon) H^2 a^2 h_{\mu\nu} + \frac{1}{2} H^2 a^2 [ (3-\epsilon) h + 2 \epsilon u^\rho u^\sigma h_{\rho\sigma} ] g_{\mu\nu} \eqend{,}
\end{split}
\end{equations}
where $\nabla^2 \equiv \nabla^\rho\nabla_\rho$, $h \equiv g^{\mu\nu} h_{\mu\nu}$ and the Hubble and deceleration parameters were defined in Eq.~\eqref{eq:slow_roll_parameters}. The perturbative expansion of $\tilde{T}_{\mu\nu}$ yields
\begin{equations}
T_{\mu\nu} &= \frac{1}{2} \left[ \dot\phi^2 - V(\phi) \right] g_{\mu\nu} + \dot\phi^2 u_\mu u_\nu \eqend{,} \\
\begin{split}
T_{\mu\nu}^{(1)} &= - 2 \dot\phi u_{(\mu} \partial_{\nu)} \phi^{(1)} + \frac{1}{2} a^2 \left[ \dot\phi^2 - V(\phi) \right] h_{\mu\nu} \\
&\quad+ \frac{1}{2} \left[ \dot\phi^2 a^2 u^\rho u^\sigma h_{\rho\sigma} + 2 \dot\phi u^\rho \partial_\rho \phi^{(1)} - V'(\phi) \phi^{(1)} \right] g_{\mu\nu} \eqend{.}
\end{split}
\end{equations}

At the background level, Einstein's equations~\eqref{eq:einstein} imply that
\begin{equations}[eq:friedmann]
- (3-2\epsilon) H^2 g_{\mu\nu} &= \frac{1}{4} \kappa^2 \left[ \dot\phi^2 - V(\phi) \right] g_{\mu\nu} \eqend{,} \\
H^2 \epsilon u_\mu u_\nu &= \frac{1}{4} \kappa^2 \dot\phi^2 u_\mu u_\nu \eqend{,}
\end{equations}
which are nothing else but Friedmann's equations relating the inflaton and its potential with the Hubble and deceleration parameters~\cite{mukhanov_feldman_brandenberger_pr_1992}. The assumption that the background spacetime satisfies Einstein's equations implies that the tensor
\begin{equation}
\tilde{E}_{\mu\nu} \equiv \tilde{G}_{\mu\nu} - \frac{\kappa^2}{2} \tilde{T}_{\mu\nu}
\end{equation}
is gauge invariant at first order in perturbation theory~\cite{stewart_walker_prsla_1974}. To obtain the constraint equations and the equations of motion for the linearised metric perturbations in the GLC coordinates, we take the first-order perturbation
\begin{equation}
\label{eq:E_mu_nu_1}
E_{\mu\nu}^{(1)} \equiv G_{\mu\nu}^{(1)} - \frac{\kappa^2}{2} T_{\mu\nu}^{(1)}
\end{equation}
and substitute the metric perturbation $h_{\mu\nu}$ with the components given in Eq.~\eqref{eq:h_mu_nu_first_order_glc}. However, since $E_{\mu\nu}^{(1)}$ is gauge invariant at linear order, we can make the computation easier by directly substituting in Eq.~\eqref{eq:E_mu_nu_1} the gauge-invariant parts of $h_{\mu\nu}$ and $\phi^{(1)}$. The gauge-invariant part of the metric perturbation in the GLC coordinates was given in Eq.~\eqref{eq:gauge_inv_metric_perturbation_glc}, and only involves $\Upsilon^{(1)}$, $\mathsf{U}_a^{(1)}$ and $\gamma_{ab}^{(1)}$.\footnote{For the definition of the gauge-invariant metric perturbation at all orders, see Eq.~\eqref{eq:inv_metric}.} The inflaton is a scalar field, and its gauge-invariant part $\Phi$ can be constructed as in Eq.~\eqref{eq:scalar_observable_expansion}. Its expansion to second order reads
\begin{splitequation}
\label{eq:Phi}
\Phi(\tilde{X}) \equiv \tilde{\phi}[x(\tilde{X})] = \phi(\tau) + \kappa \Phi^{(1)}(\tilde{X}) + \kappa^2 \Phi^{(2)}(\tilde{X}) + \bigo{\kappa^3} \eqend{,}
\end{splitequation}
where the first-order gauge-invariant correction $\Phi^{(1)}$ is related to the gauge-dependent inflaton perturbation $\phi^{(1)}$~\eqref{eq:perturbed_inflaton} according to
\begin{equation}
\label{eq:Phi_1}
\Phi^{(1)} = \phi^{(1)} - \dot\phi \tau_{(1)} \eqend{.}
\end{equation}

After performing these substitutions, we can write down the constraint equations and the equations of motion for the perturbations. The constraint equations are obtained by computing the contraction
\begin{equation}
\Tilde{E}_\mu \equiv \tilde{E}_{\mu\nu} \tilde{u}^\nu = \kappa E_{\mu\nu}^{(1)} u^\nu + \bigo{\kappa^2} \equiv \kappa E_\mu^{(1)} + \bigo{\kappa^2} \eqend{,}
\end{equation}
where the perturbed four-velocity $\tilde{u}^\mu$ was defined in Eq.~\eqref{eq:u_mu_tilde}. The resulting equations are
\begin{splitequation}
\label{eq:E_tau}
0 = E_\tau^{(1)} &= - \frac{1}{2 a^3 r^2} \left[ 4 a r (1 - H a r) \partial_\tau + 4 \left( - 1 + H a r + H^2 a^2 r^2 \right) + \laplace_\Omega \right] \Upsilon^{(1)} \\
&\quad- \frac{1}{2} \kappa^2 \left[ \dot{\phi} \partial_\tau + \frac{1}{2} V'(\phi) \right] \Phi^{(1)} + \frac{1}{2 a^3 r^2} \left( \partial_\tau + 2 H - \frac{2}{a r} \right) D_a U^a_{(1)} + \frac{1}{2 a^4 r^4} D_a D_b \gamma_{(1)}^{ab} \\
&\quad+ \frac{1}{2 a^3 r^2} \left[ \left( \partial_w + 2 H a - \frac{1}{r} \right) \partial_\tau - \frac{1}{a r^2} \laplace_\Omega - \frac{1}{a r^2} \left( 1 - 2 H a r + 4 H^2 a^2 r^2 \right) \right] \gamma^{(1)} \eqend{,}
\end{splitequation}
\begin{splitequation}
\label{eq:E_w}
0 = E_w^{(1)} &= - \frac{1}{2 a^2 r^2} \left[ - 4 a r \left( \partial_\tau + a^{-1} \partial_w - H \right) + \laplace_\Omega \right] \Upsilon^{(1)} - \frac{1}{2} \kappa^2 \dot \phi \partial_w \Phi^{(1)} \\
&\quad- \frac{1}{2 a^2 r^2} \left( \partial_\tau + \frac{2}{a} \partial_w - 2 H \right) D_a U^a_{(1)} - \frac{1}{2 a^2 r^2} \bigg[ \partial_\tau \partial_w + a^{-1} \partial_w^2 \\
&\quad- \frac{1}{r} \partial_\tau - \left( 2 H + \frac{1}{a r} \right) \partial_w + \frac{2 H}{r} \bigg] \gamma^{(1)} \eqend{,}
\end{splitequation}
and for the two angular components
\begin{splitequation}
\label{eq:E_a}
0 = E_b^{(1)} &= - \frac{1}{a} \partial_b \left( \partial_\tau + \frac{1}{2 a} \partial_w - H \right) \Upsilon^{(1)} - \frac{1}{2} \kappa^2 \dot{\phi} \partial_b \Phi^{(1)} - \frac{1}{2 a^2} \bigg[ \partial_\tau \partial_w - 2 \left( H - \frac{1}{a r} \right) \partial_w \\
&\quad + \frac{2}{r} \partial_\tau - \frac{4 H}{r} + \frac{1}{a r^2} \bigg] U_b^{(1)} - \frac{1}{2 a^3 r^2} \left( D_b D_a U_{(1)}^a - D^a D_a U^{(1)}_b \right) \\
&\quad + \frac{1}{2 a^2 r^2} \left( \partial_\tau + \frac{1}{a} \partial_w - 2 H \right) \left( D^a \gamma^{(1)}_{ab} - D_b \gamma^{(1)} \right) \eqend{.}
\end{splitequation}
In the expressions above, $D_a$ is the covariant derivative of the metric $s_{ab}$ on $\mathbb{S}^2$, see Eq.~\eqref{eq:flrw_metric_conformal}, $\laplace_\Omega f \equiv s^{ab} D_a D_b f$, the trace $\gamma^{(1)} \equiv s^{ab}\gamma^{(1)}_{ab}$ and we recall that $r = w - \eta$, which we used to shorten the expressions. The other six equations are dynamical, and read
\begin{splitequation}
\label{eq:E_ww}
0 = E_{ww}^{(1)} &= - \frac{1}{2 r^2} \left[ \partial_\tau^2 + \frac{2}{a} \partial_\tau \partial_w + \frac{1}{a^2} \partial_w^2 - H \partial_\tau - \left( \frac{2 H}{a} + \frac{1}{a^2 r} \right) \partial_w - 2 H^2 (1-\epsilon) + \frac{1}{a^2 r^2} \right] \gamma^{(1)} \\
&\quad- \frac{1}{2 r^2} \frac{1}{a^2 r^2} \left( D_a D_b \gamma_{(1)}^{ab} - \laplace_\Omega \gamma^{(1)} \right) - \frac{1}{a r^2} \left( \partial_\tau + \frac{1}{a} \partial_w - \frac{1}{a r} \right) D_a U_{(1)}^a \\
&\quad+ \frac{2}{r} \left( \partial_\tau + \frac{1}{a} \partial_w + H - \frac{1}{a r} \right) \Upsilon^{(1)} - \frac{1}{2} \kappa^2 a^2 \dot\phi \left[ \partial_\tau + \frac{1}{a} \partial_w - \frac{1}{2 \dot\phi} V'(\phi) \right] \Phi^{(1)} \eqend{,}
\end{splitequation}
\begin{splitequation}
\label{eq:E_wb}
0 = E_{wb}^{(1)} &= - \frac{1}{2} \left[ \partial_\tau^2 + \frac{1}{a} \partial_\tau \partial_w - \left( H - \frac{2}{a r} \right) \partial_\tau - \frac{2}{a} \left( H - \frac{1}{a r} \right) \partial_w - 2 H^2 (1 - \epsilon) - \frac{1}{a^2 r^2} \right] U^{(1)}_b \\
&\quad+ \frac{1}{2 a^2 r^2} \left( D^a D_a U^{(1)}_b - D_b D_a U_{(1)}^a \right) + \frac{1}{2 a^2 r^2} \left( \partial_w - \frac{2}{r} \right) \left( D^a \gamma_{ab}^{(1)} - D_b \gamma^{(1)} \right) \\
&\quad- \frac{1}{2} D_b \left( \partial_\tau + \frac{1}{a} \partial_w + H - \frac{2}{a r} \right) \Upsilon^{(1)}
\end{splitequation}
and
\begin{splitequation}
\label{eq:E_ab}
0 = E_{ab}^{(1)} &= \frac{1}{2} \left[ \partial_\tau^2 + \frac{2}{a} \partial_\tau \partial_w - H \partial_\tau - \frac{2}{a} \left( H - \frac{1}{a r} \right) \partial_w - 2 H^2 (1-\epsilon) - \frac{2}{a^2 r^2} \right] \gamma^{(1)}_{ab} \\
&\quad - \frac{1}{a} D_a D_b \Upsilon^{(1)} + \frac{1}{a} \partial_\tau D_{(a} U^{(1)}_{b)} + \frac{1}{2} \gamma_{ab} \kappa^2 a^2 r^2 \dot\phi \left[ \partial_\tau + \frac{1}{a} \partial_w - \frac{1}{2 \dot\phi} V'(\phi) \right] \Phi^{(1)} \\
&\quad+ a r^2 \gamma_{ab} \left[ \partial_\tau^2 + \frac{1}{a} \partial_\tau \partial_w + \left( H - \frac{1}{a r} \right) \partial_\tau - \frac{H}{a} \partial_w - (2-\epsilon) H^2 - \frac{H}{a r} \right] \Upsilon^{(1)}\eqend{.}
\end{splitequation}
Note that there is no second time derivative for $\Phi^{(1)}$ in Eqs.~\eqref{eq:E_ww}--\eqref{eq:E_ab}. Furthermore, the combination of the trace of Eq.~\eqref{eq:E_ab} with Eq.~\eqref{eq:E_ww} results in
\begin{splitequation}
0 &= \gamma^{ab} E_{ab}^{(1)} - r^2 E_{ww}^{(1)} \\
&= \frac{1}{2 a^2} \left( \partial_w^2 - \frac{3}{r} \partial_w + \frac{3}{r^2} \right) \gamma^{(1)} + \frac{1}{2 a^2 r^2} \left( D_a D_b \gamma_{(1)}^{ab} - \laplace_\Omega \gamma^{(1)} \right) + \frac{1}{a^2} \left( \partial_w - \frac{1}{r} \right) D_a U_{(1)}^a \\
&\quad- 2 a r^2 \left[ \partial_\tau^2 + \frac{1}{a} \partial_\tau \partial_w + H \partial_\tau - \frac{H}{a} \partial_w + \frac{1}{a^2 r} \partial_w - (2-\epsilon) H^2 - \frac{1}{a^2 r^2} \right] \Upsilon^{(1)} \\
&\quad + \frac{1}{a} \laplace_\Omega \Upsilon^{(1)} - \frac{1}{2} \kappa^2 a^2 r^2 \dot\phi \left[ \partial_\tau + \frac{1}{a} \partial_w - \frac{1}{2 \dot\phi} V'(\phi) \right] \Phi^{(1)} \eqend{,}
\end{splitequation}
the equation of motion for $\Upsilon^{(1)}$.

The equation of motion for $\Phi^{(1)}$ is obtained from the vanishing divergence of the stress tensor. Using Friedmann's equations~\eqref{eq:friedmann} (and their time derivatives), we have
\begin{equation}
\tilde{\nabla}^\mu \tilde{T}_{\mu\nu} = - u_\nu \left[ \tilde{\nabla}^2 \tilde\phi - \frac{1}{2} V'(\tilde{\phi}) \right] \dot\phi + \bigo{\kappa^2} \eqend{.}
\end{equation}
Thus, in the perturbative expansion of the contraction
\begin{equation}
\tilde{F}_\Phi \equiv - \frac{1}{4 H^2 \epsilon} \tilde{u}^\nu \tilde{\nabla}^\mu \tilde{T}_{\mu\nu} = \kappa F_\Phi^{(1)} + \bigo{\kappa^2} \eqend{,}
\end{equation}
the scalar $F^{(1)}_\Phi$ is gauge invariant at linear order. This yields
\begin{splitequation}\label{eq:eom_Phi_1}
0 = F_\Phi^{(1)} &= \bigg\{ \partial_\tau^2 + \frac{2}{a} \partial_\tau \partial_w + \left( 3 - 2 \epsilon + 2 \delta \right) H \partial_\tau + \frac{2}{a} \left[ (1 - \epsilon + \delta) H - \frac{1}{a r} \right] \partial_w \\
&\quad+ 3 H^2 \epsilon - \frac{1}{a^2 r^2} \laplace_\Omega \bigg\} \left( \frac{\Phi^{(1)}}{\dot{\phi}} \right) + \frac{1}{a} \left( \partial_\tau - H - \frac{2}{a r} \right) \Upsilon^{(1)}\\
&\quad+ \frac{1}{2a^2 r^2} \left( \partial_\tau + a^{-1} \partial_w - 2 H \right) \gamma^{(1)} + \frac{1}{a^3 r^2} D_a U_{(1)}^a \eqend{,}
\end{splitequation}
and we recall that the slow-roll parameters $H$, $\epsilon$ and $\delta$ were defined in Eq.~\eqref{eq:slow_roll_parameters}. Moreover, Eq.~\eqref{eq:eom_Phi_1} can be expressed in terms of Eqs.~\eqref{eq:E_tau}--\eqref{eq:E_ab} as
\begin{splitequation}
F^{(1)}_\Phi &= - \frac{1}{2 H^2 \epsilon} \left[ \left( \partial_\tau + a^{-1} \partial_w + 3 H \right) E_\tau^{(1)} + a^{-1} \left( \partial_\tau + 2 H \right) E_w^{(1)} \right] \\
&\quad+ \frac{1}{2 a^2 H^2 \epsilon r^2} \left( D^a E_a^{(1)} - H r^2 E_{ww}^{(1)} - H \gamma^{ab} E_{ab}^{(1)} + 2 r E_w^{(1)} \right) \eqend{.}
\end{splitequation}
This last result comes as a consequence of the contracted Bianchi identity, and serves as a check of our expressions.

In contrast with the usual 3+1 scalar-vector-tensor (SVT) decomposition~\cite{lifshitz_grg_2017} of the metric perturbation, we see from Eqs.~\eqref{eq:E_tau}--\eqref{eq:E_ab} and~\eqref{eq:eom_Phi_1} that the GLC decomposition of $h_{\mu\nu}$ results in a system of coupled equations for the gauge-independent variables. An important difference between the SVT and GLC decompositions is that the SVT decomposition takes full advantage of the background symmetries by employing spatial scalars, vectors and tensors, which transform irreducibly under rotations and translations of the spatial hypersurfaces and thus decouple. On the other hand, the GLC decomposition expresses the metric perturbation in terms of the scalar $\Upsilon^{(1)}$ and the quantities $U^{(1)}_a$ and $\gamma^{(1)}_{ab}$, which transform only under rotations and are furthermore reducible, containing the scalars $D^a U^{(1)}_a$, $\gamma^{(1)}$ and $D^a D^b \gamma^{(1)}_{ab}$. Therefore, the linearised Einstein's equations remain coupled. In the next section, we compare and relate these two decompositions for $h_{\mu\nu}$ in more detail.

\subsection{Comparison with the SVT decomposition}

It is interesting to compare the GLC decomposition of the metric perturbation $h_{\mu\nu}$ given in Eq.~\eqref{eq:h_mu_nu_first_order_glc} with the more familiar one in the conformally flat coordinates~\eqref{eq:flrw_conformallyflat}. Thus, we decomposed $h_{\mu\nu}$ according to its transformation under rotations and translations of the spatial hypersurfaces. In Cartesian coordinates, this decomposition can be written as~\cite{froeb_jcap_2014}
\begin{equations}[eq:h_mu_nu_phi_first_order_conformal]
h_{\eta\eta} & = S + 2 X_\eta' + 2 H a X_\eta \eqend{,} \\
h_{\eta i} & = V_i + X_i' + \partial_i X_\eta \eqend{,} \\
h_{ij} & = H_{ij} + \delta_{ij} \Sigma + 2 \partial_{(i} X_{j)} - 2 H a \delta_{ij} X_\eta \eqend{,} \\
\phi^{(1)} & = \Psi - \phi' X_\eta \eqend{,}
\end{equations}
where Latin indices denote spatial components. In the expressions above, $S$, $\Sigma$ and $\Psi$ are scalars, $V_i$ is a transverse vector ($\partial^i V_i \equiv \delta^{ij} \partial_i V_j = 0$), $H_{ij}$ is a symmetric transverse traceless tensor ($\partial^i H_{ij} = 0 = \delta^{ij} H_{ij}$) and a prime denotes the derivative with respect to the conformal time $\eta$. Under a gauge transformation, the metric and inflaton perturbations in the conformal coordinates transform as
\begin{equation}
a^2 \delta_\xi h_{\mu\nu} = \partial_\mu \xi_\nu + \partial_\nu \xi_\mu - 2 H a \left( \delta^\eta_\mu \xi_\nu + \delta^\eta_\nu \xi_\mu + \eta_{\mu\nu} \xi_\eta \right) \quad\text{and}\quad \delta_\xi \phi^{(1)} = - \phi' \xi^\eta \eqend{,}
\end{equation}
respectively, where $\eta_{\mu\nu}$ is the Minkowski metric. This transformation leaves $S$, $\Sigma$, $\Psi$, $V_i$ and $H_{ij}$ invariant, whereas $X_\mu$ changes as $\delta_\xi X_\mu = a^{-2}\xi_\mu$.

We now wish to relate these invariants to $\Upsilon^{(1)}$, $U_a^{(1)}$, $\Phi^{(1)}$ and $\gamma_{ab}^{(1)}$, the invariants appearing in the GLC decomposition of the metric and inflaton perturbations. To do this, we first express $S$, $\Sigma$, $\Psi$, $V_i$ and $H_{ij}$ in terms of $h_{\mu\nu}$ and $\phi^{(1)}$ in an arbitrary gauge by inverting Eq.~\eqref{eq:h_mu_nu_phi_first_order_conformal}. The result for an $n$-dimensional FLRW spacetime is
\begin{equations}[eq:inv_variables_conformal_coord]
\laplace^2 \Psi &= \laplace \left( \laplace \phi^{(1)} + \phi' \partial^i h_{\eta i} \right) - \frac{\phi'}{2 (n-2)} \partial_\eta \left[ (n-1) \partial^i \partial^j h_{ij} - \laplace\left(\delta^{ij} h_{ij}\right) \right] \eqend{,} \\
\laplace^2 S &= \laplace \left[ \laplace h_{\eta\eta} - \frac{2}{a} \partial_\eta \left( a \partial^i h_{\eta i} \right) \right] + \frac{1}{(n-2) a} \partial_\eta \left\{ a \partial_\eta \left[ (n-1) \partial^i \partial^j h_{ij} - \laplace \left( \delta^{ij} h_{ij} \right) \right] \right\} \eqend{,} \\
\laplace^2 \Sigma &= \frac{1}{n-2} \left( \laplace + H a \partial_\eta \right) \Pi^{ij} h_{ij} + H a \left( 2 \laplace \partial^i h_{\eta i} - \partial_\eta \partial^i \partial^j h_{ij} \right) \eqend{,} \\
\laplace^2 V_i &= \Pi_i{}^j \left( \laplace h_{\eta j} - \partial_\eta \partial^k h_{jk} \right) \eqend{,} \\
\laplace^2 H_{ij} &= \Pi_i{}^k \Pi_j{}^l h_{kl} - \frac{1}{n-2} \Pi_{ij} \Pi^{kl} h_{kl} \eqend{,}
\end{equations}
where $\laplace$ denotes the usual Euclidean Laplace operator and we have defined the projection
\begin{equation}
\Pi_i{}^j \equiv \laplace \delta_i^j - \partial_i \partial^j \eqend{.}
\end{equation}
In terms of the decomposition presented above, the Sasaki--Mukhanov variable~\cite{mukhanov_feldman_brandenberger_pr_1992} is defined as
\begin{equation}
Q \equiv \frac{2 H a}{\phi'}\Psi - \Sigma \eqend{,}
\end{equation}
and we can use Eqs.~\eqref{eq:inv_variables_conformal_coord} to relate it to the metric and inflaton perturbations, which gives
\begin{equation}
\laplace Q = \frac{2 H a}{\phi'} \laplace \phi^{(1)} - \frac{1}{n-2} \laplace \left( \delta^{ij} h_{ij} \right) + \frac{1}{n-2} \partial^i \partial^j h_{ij} \eqend{.}
\end{equation}

Next, we have to transform these expressions into the GLC coordinates. To do this, we pass to spherical coordinates and then transform the coordinates $\eta$ and $r$ to $\tau$ and $w$, see Eqs.~\eqref{eq:w_coord_flrw} and~\eqref{eq:flrw_metric_glc}. After performing these transformations in Eqs.~\eqref{eq:inv_variables_conformal_coord}, we obtain the following equations for the scalars $S$, $\Sigma$ and $\Psi$ in (for simplicity) $n = 4$ dimensions:
\begin{splitequation}
\label{eq:S_to_glc}
\laplace^2 S &= - 2 \partial_\tau \laplace \left[\left( \partial_r + \frac{2}{r} \right) \Upsilon^{(1)} - \frac{1}{a r^2} D_a U_{(1)}^a \right] + \partial_\tau \bigg\{ a^2 \partial_\tau \bigg[ \frac{2}{a} \left( \partial_r^2 + \frac{5}{r} \partial_r + \frac{3}{r^2} \right) \Upsilon^{(1)}\\
&\quad- \frac{1}{a r^2} \laplace_\Omega \Upsilon^{(1)} - \frac{3}{a^2 r^2} \left( \partial_r + \frac{1}{r} \right) D_a U_{(1)}^a - \frac{1}{2 a^2 r^2} \left( \partial_r^2 + \frac{1}{r} \partial_r - \frac{1}{r^2} + \laplace_\Omega \right) \gamma^{(1)}\\
&\quad+ \frac{3}{2 a^2 r^4} D^a D^b \gamma_{ab}^{(1)} \bigg] \bigg\} \eqend{,}
\end{splitequation}
\begin{splitequation}
\label{eq:Sigma_to_glc}
\laplace^2 \Sigma &= \laplace \bigg\{\left[ 2 H \left( \partial_r + \frac{2}{r} \right) - \frac{2}{a r} \left( \partial_r + \frac{1}{r} \right) + \frac{1}{a r^2} \laplace_\Omega \right]\Upsilon^{(1)} + \frac{1}{a^2r^2} \left( \partial_r + \frac{1}{r} - 2 H a \right) D_a U_{(1)}^a\\
&\quad- \frac{1}{2a^2r^4} \left( D_a D_b \gamma_{(1)}^{ab} - \laplace_\Omega \gamma^{(1)} \right) + \frac{1}{2a^2r^2} \left( \partial_r^2 - \frac{1}{r} \partial_r + \frac{1}{r^2} \right) \gamma^{(1)} \bigg\} \\
&\quad+ H a^2 \partial_\tau \bigg[ - \frac{2}{a} \left( \partial_r^2 + \frac{5}{r} \partial_r + \frac{3}{r^2} \right) \Upsilon^{(1)} + \frac{1}{a r^2} \laplace_\Omega \Upsilon^{(1)} + \frac{3}{a^2 r^2} \left( \partial_r + \frac{1}{r} \right) D_a U_{(1)}^a \\
&\quad+ \frac{1}{2 a^2 r^2} \left( \partial_r^2 + \frac{1}{r} \partial_r - \frac{1}{r^2} \right) \gamma^{(1)} + \frac{1}{2 a^2 r^4} \laplace_\Omega \gamma^{(1)} - \frac{3}{2 a^2 r^4} D_a D_b \gamma^{ab}_{(1)} \bigg] \raisetag{2em}
\end{splitequation}
and
\begin{splitequation}
\label{eq:Psi_to_glc}
\laplace^2 \Psi &= \laplace^2 \Phi^{(1)} + \dot{\phi} \laplace \left[ \left( \partial_r + \frac{2}{r} \right) \Upsilon^{(1)} - \frac{1}{a r^2} D_a U_{(1)}^a \right] - \frac{a^2}{4} \dot{\phi} \partial_\tau \bigg[ \frac{4}{a} \left( \partial_r^2 + \frac{5}{r} \partial_r + \frac{3}{r^2} \right) \Upsilon^{(1)}\\
&\quad- \frac{2}{a r^2} \laplace_\Omega \Upsilon^{(1)} - \frac{6}{a^2 r^2} \left( \partial_r + \frac{1}{r} \right) D_a U_{(1)}^a - \frac{1}{a^2 r^2} \left( \partial_r^2 + \frac{1}{r} \partial_r - \frac{1}{r^2} \right) \gamma^{(1)}\\
&\quad- \frac{1}{a^2 r^4} \laplace_\Omega \gamma^{(1)} + \frac{3}{a^2 r^4} D_a D_b \gamma^{ab}_{(1)} \bigg] \eqend{,} \raisetag{2em}
\end{splitequation}
and we recall that $r = w - \eta(\tau)$. In passing we note that the gauge-invariant inflaton perturbation in the GLC coordinates, $\Phi^{(1)}$, differs from its counterpart $\Psi$ in the $3 + 1$ decomposition. Furthermore, the equation relating the Sasaki--Mukhanov variable to the metric perturbations in the GLC gauge is
\begin{splitequation}
\laplace Q &= \frac{2 H}{\dot{\phi}} \laplace \Phi^{(1)} + \frac{2}{a r} \left( \partial_r + \frac{1}{r} \right) \Upsilon^{(1)} - \frac{1}{a r^2} \laplace_\Omega \Upsilon^{(1)} - \frac{1}{a^2 r^2} \left( \partial_r + \frac{1}{r} \right) D_a U_{(1)}^a \\
&\quad+ \frac{1}{2 a^2 r^4} \left( D_a D_b \gamma_{(1)}^{ab} - \laplace_\Omega \gamma^{(1)} \right) - \frac{1}{2 a^2 r^2} \left( \partial_r^2 - \frac{1}{r} \partial_r + \frac{1}{r^2} \right) \gamma^{(1)} \eqend{,}
\end{splitequation}
and we recall that $\gamma^{(1)} \equiv s^{ab}\gamma^{(1)}_{ab}$.

We can now use Eqs.~\eqref{eq:S_to_glc}--\eqref{eq:Psi_to_glc} to compare the constraint equations for the perturbations in the GLC coordinates to the ones in the conformal coordinates. For example, for the scalar sector, the invariant combination
\begin{equation}
F \equiv (3-\epsilon) H^2 a^2 S + 3 H a \partial_\eta \Sigma - \laplace \Sigma - \frac{\kappa^2}{4} \left[ 2 \phi' \partial_\eta \Psi + a^2 V'(\phi) \Psi \right]
\end{equation}
vanishes when Einstein's equations are satisfied. Indeed, we compute that
\begin{splitequation}
\label{eq:F_eq}
\laplace^2 F &= - \laplace^2 \left[ \laplace \Sigma + H a^2 \epsilon \partial_\tau Q \right] + (3-\epsilon) H^2 a^2 \laplace^2 \left[ S + \frac{1}{H} \partial_\tau \Sigma + \epsilon (Q + \Sigma) \right] \\
&= a^2 \laplace^2 \left( E_\tau^{(1)} + a^{-1} E_w^{(1)} \right) \eqend{,}
\end{splitequation}
for which we have used that the term $\laplace \Sigma + H a^2 \epsilon \partial_\tau Q$ satisfies
\begin{equation}
\label{eq:Sigma_Q_combination_eq}
\laplace \left[ \laplace \Sigma + H a^2 \epsilon \partial_\tau Q \right] = - a^2 \laplace \left( E_\tau^{(1)} + a^{-1} E_w^{(1)} \right) - (3-\epsilon) H a^2 \left( \frac{1}{r^2} D^a E_a^{(1)} + \partial_w E_w^{(1)} + \frac{2}{r} E_w^{(1)} \right) \eqend{.}
\end{equation}
Hence, both Eqs.~\eqref{eq:F_eq} and~\eqref{eq:Sigma_Q_combination_eq} vanish when the constraint equations~\eqref{eq:E_tau}--\eqref{eq:E_a} hold, assuming a sufficiently fast fall-off of perturbations at spatial infinity. Similar checks can be made for the vector constraint.

Lastly, we would like to relate the equation of motion for $Q$ to Einstein's equations in the GLC coordinates. This equation reads~\cite{mukhanov_feldman_brandenberger_pr_1992}
\begin{equation}
\label{eq:eom_Q}
F_Q \equiv \partial_\tau^2 Q + \left( 3 + 2 \delta \right) H \partial_\tau Q - a^{-2} \laplace Q = 0 \eqend{.}
\end{equation}
By taking the Laplacian on both sides of the equation above and rearranging the terms, we obtain
\begin{splitequation}
\laplace F_Q &= 2 H \laplace F_\Phi^{(1)} + \frac{2}{a^2 r^2} \left( \partial_w + \frac{1}{r} \right) D^a E_{wa}^{(1)} + \frac{1}{a^2 r^4} D^a D^b E_{ab}^{(1)} - \frac{1}{a^2 r^3} \left( \partial_w - \frac{1}{r} \right) \gamma^{ab} E_{ab}^{(1)} \\
&\quad+ \frac{1}{a^2 r^2} \partial_w^2 \left( r^2 E_{ww}^{(1)} \right) - \laplace \left( E_\tau^{(1)} + a^{-1} E_w^{(1)} \right) + 2H\delta \left( \partial_w E_w^{(1)} + \frac{2}{r} E_w^{(1)} + \frac{1}{r^2} D^a E_a^{(1)} \right) \eqend{.}
\end{splitequation}
That is, if the dynamical and constraint equations resulting from the linearised Einstein's equations are fulfilled, then the Sasaki--Mukhanov variable $Q$ satisfies its well-known equation of motion~\eqref{eq:eom_Q}. We note that this is not a consequence of expressing $Q$ in the GLC coordinates. In fact, also in the conformally flat coordinates Eq.~\eqref{eq:eom_Q} only holds when the constraint equations are satisfied, see for example Refs.~\cite{mukhanov_feldman_brandenberger_pr_1992,froeb_jcap_2014}. Thus, as in the standard treatment of the Sasaki--Mukhanov variable, also here we do not expect any complications to arise from the off-shell terms in the quantisation of the system.

\section{Gauge-invariant Hubble rate}                                                                                                          %
\label{sec:hubble_parameter}                                                                                                                        %

In this section we apply the method to construct gauge-invariant observables described in Sec.~\ref{sec:gauge_invariant_obs} to obtain an invariant measure of the local expansion rate of the spacetime. This observable (which we shall refer to as the local Hubble rate, for short) can be defined as the expansion rate of the spatial section in any time foliation of the spacetime. Here we will explore two alternative definitions of the local Hubble rate: one is to use the observer's spatial section to define this observable, and the other is to use the level hypersurfaces of the inflaton field. Our aim here is to find explicit formulas for both these rates in terms of the gauge-invariant perturbations of the metric and inflaton field, up to second order in perturbation theory.

Let us start by analysing the local Hubble rate defined by the observer's spatial section. In this case, the normal vector field in the perturbed spacetime is the four-velocity $\tilde{u}^\mu$, and the (gauge-dependent) Hubble rate is proportional to its divergence~\cite{geshnizjani_brandenberger_prd_2002}:
\begin{equation}
\label{eq:H_u_tilde}
\tilde{H}_u(x) \equiv \frac{1}{n-1} \tilde{\nabla}_\mu \tilde{u}^\mu(x) \eqend{.}
\end{equation}
To make this scalar gauge invariant, we have to transform from the background GLC coordinates $x^\mu = (\tau, w, \theta^a)$ to the field-dependent GLC coordinates $\tilde{X}^{(\mu)} = (\tilde{\tau}, \tilde{w}, \tilde{\theta}^a)$. The result is
\begin{equation}
\label{eq:H_U_inv_def}
\mathcal{H}_u(\tilde{X}) \equiv \tilde{H}_u[x(\tilde{X})] \eqend{.}
\end{equation}
The perturbative expansion for $\mathcal{H}_u$ can then be obtained as in Eq.~\eqref{eq:scalar_observable_expansion}, \ie, by combining the expansions of $\tilde{H}_u$ and the inverse of $\tilde{X}^{(\mu)}$ up to second order in $h_{\mu\nu}$ and $\phi^{(1)}$, and then rearranging the terms, verifying that all gauge-dependent terms cancel.

However, since here we are mainly interested in the invariant final expression for the Hubble rate, it is more convenient to transform the right-hand side of Eq.~\eqref{eq:H_u_tilde} to the coordinates $\tilde{X}^{(\mu)}$ before performing the perturbative expansion. Using the gauge-invariant covariant derivative $\bnabla_\mu$ defined in Eq.~\eqref{eq:inv_derivative} and the invariant four-velocity
\begin{equation}
\label{eq:U_inv_def}
\mathcal{U}^\mu(\tilde{X}) \equiv \frac{\partial \tilde{X}^{(\mu)}}{\partial x^\nu} \tilde{u}^\nu[x(\tilde{X})] \eqend{,}
\end{equation}
we can write Eq.~\eqref{eq:H_U_inv_def} as
\begin{equation}
\label{eq:H_U_inv_div}
\mathcal{H}_u(\tilde{X}) = \frac{1}{n-1} \bnabla_\mu \mathcal{U}^\mu(\tilde{X}) \eqend{.}
\end{equation}
To find the perturbative expansion for Eq.~\eqref{eq:H_U_inv_div}, we first expand the derivative operator. This yields
\begin{splitequation}
\label{eq:div_U_inv_expansion_1}
\bnabla_\mu \mathcal{U}^\mu &= \nabla_\mu \mathcal{U}^\mu + \mathcal{C}_{\rho\mu}^\rho \mathcal{U}^\mu \\
&= \nabla_\mu \mathcal{U}^\mu + \frac{\kappa}{2} \mathcal{U}^\mu \nabla_\mu \left( g^{\alpha\beta} \mathcal{H}_{\alpha\beta} \right) - \frac{\kappa^2}{4} \mathcal{U}^\mu \nabla_\mu \left( \mathcal{H}^{\alpha\beta} \mathcal{H}_{\alpha\beta} \right) + \bigo{\kappa^3} \eqend{,}
\end{splitequation}
where that the invariant tensor $\mathcal{C}^\rho_{\mu\nu}$ and the invariant metric perturbation $\mathcal{H}_{\mu\nu}$ were defined in Eqs.~\eqref{eq:C} and~\eqref{eq:inv_metric_perturbation}, respectively.

The perturbative expansion of the invariant four-velocity $\mathcal{U}^\mu$ can be easily obtained by noticing that using Eq.~\eqref{eq:U_inv_def} and the definition~\eqref{eq:u_mu_tilde} of $\tilde{u}^\mu$, we obtain
\begin{equation}
\mathcal{U}^\mu = - \frac{\partial \tilde{X}^{(\mu)}}{\partial x^\nu} \frac{\partial \tilde{\tau}}{\partial x^\rho} \tilde{g}^{\nu\rho} = - \mathcal{G}^{\mu\tau} \eqend{,}
\end{equation}
the inverse invariant metric defined in Eq.~\eqref{eq:inverse_inv_metric}. Hence, up to second order, we have
\begin{equation}
\label{eq:U_inv_expansion}
\mathcal{U}^\mu = u^\mu + \kappa \mathcal{H}^{\mu\tau} - \kappa^2 \mathcal{H}^{\mu\nu} \mathcal{H}_\nu{}^\tau + \bigo{\kappa^3} \eqend{.}
\end{equation}
Combining this expression with Eq.~\eqref{eq:div_U_inv_expansion_1}, we obtain
\begin{splitequation}
\label{eq:div_U_inv_expansion_2}
\bnabla_\mu \mathcal{U}^\mu &= \nabla_\mu u^\mu + \frac{\kappa}{2} \left[ u^\mu \nabla_\mu \left( g^{\alpha\beta} \mathcal{H}_{\alpha\beta} \right) + 2 \nabla_\mu \mathcal{H}^{\mu\tau} \right] \\
&\quad- \frac{\kappa^2}{4} \left[ u^\mu \nabla_\mu \left( \mathcal{H}^{\alpha\beta} \mathcal{H}_{\alpha\beta} \right) - 2 \mathcal{H}^{\tau\mu} \nabla_\mu \left( g^{\alpha\beta} \mathcal{H}_{\alpha\beta} \right) + 4 \nabla_\mu \left( \mathcal{H}^{\mu\nu} \mathcal{H}_\nu{}^\tau \right) \right] + \bigo{\kappa^3} \eqend{.}
\end{splitequation}

We still need to express the components of the invariant metric perturbation in terms of the expansion of $\tilde{\Upsilon}$, $\tilde{\mathsf{U}}^a$ and $\tilde{\gamma}_{ab}$. From Eq.~\eqref{eq:inv_metric_perturbation}, we see that the expressions for the components of $\mathcal{H}_{\mu\nu}$ can be easily read off from the invariant metric, see Eqs.~\eqref{eq:glc_metric_matrix_form} and~\eqref{eq:glc_metric_perturbed}. Up to second order, they are
\begin{equations}[eq:H_covariant_components]
\mathcal{H}_{\tau\tau} &= \mathcal{H}_{\tau a} = 0 \eqend{,} \\
\mathcal{H}_{\tau w} &= - \Upsilon^{(1)} - \kappa \Upsilon^{(2)} \eqend{,} \\
\mathcal{H}_{ww} &= 2 a \Upsilon^{(1)} + \kappa \left[ \left( \Upsilon^{(1)} \right)^2 + 2 a \Upsilon^{(2)} + \mathsf{U}_a^{(1)} \mathsf{U}_{(1)}^a \right] \eqend{,} \\
\mathcal{H}_{wa} &= - \mathsf{U}^{(1)}_a - \kappa \left[ \mathsf{U}^{(2)}_a + \gamma^{(1)}_{ab} \mathsf{U}_{(1)}^b \right] \eqend{,} \\
\mathcal{H}_{ab} &= \gamma^{(1)}_{ab} + \kappa \gamma^{(2)}_{ab} \eqend{.}
\end{equations}
From this, we obtain the invariant metric perturbation components with both indices raised with the background metric $g^{\mu\nu}$~\eqref{eq:flrw_metric_glc}, which read
\begin{equations}[eq:H_contravariant_components]
\mathcal{H}^{\tau\tau} & = \frac{\kappa}{a^2} \left[ \left( \Upsilon^{(1)} \right)^2 + \mathsf{U}^{(1)}_a \mathsf{U}_{(1)}^a \right] \eqend{,} \\
\mathcal{H}^{\tau w} & = - \frac{1}{a^2} \left( \Upsilon^{(1)} + \kappa \Upsilon^{(2)} \right) \eqend{,} \\
\mathcal{H}^{\tau a} & = \frac{1}{a} \left[ \mathsf{U}_{(1)}^a + \kappa \left( \mathsf{U}_{(2)}^a + \gamma_{(1)}^{ab} \mathsf{U}^{(1)}_b \right) \right] \eqend{,} \\
\mathcal{H}^{ww} &= \mathcal{H}^{wa} = 0 \eqend{,} \\
\mathcal{H}^{ab} &= \gamma_{(1)}^{ab} + \kappa \gamma_{(2)}^{ab} \eqend{.}
\end{equations}
Then, by substituting Eqs.~\eqref{eq:H_covariant_components} and~\eqref{eq:H_contravariant_components} into Eq.~\eqref{eq:div_U_inv_expansion_2}, we obtain the final form for the invariant Hubble rate~\eqref{eq:H_U_inv_def}:
\begin{splitequation}
\label{eq:H_u_final}
\mathcal{H}_u &= H - \frac{\kappa}{2} \left[ \frac{1}{(n-1)} \nabla_\mu \left( 2 k^\mu \Upsilon_{(1)} - 2 a^{-1} U_{(1)}^\mu - u^\mu \gamma^{ab} \gamma^{(1)}_{ab} \right) + H a^{-1} \left( 2 \Upsilon^{(1)} + a \gamma^{ab} \gamma^{(1)}_{ab} \right) \right] \\
&\quad- \frac{\kappa^2}{4 (n-1)} \bigg\{ 4 \nabla_\mu \left[ k^\mu \left[ \Upsilon^{(2)} - a^{-1} \left( \Upsilon^{(1)} \right)^2 \right] \right] + 4 (n-1) H a^{-1} \left[ \Upsilon^{(2)} - a^{-1} \left( \Upsilon^{(1)} \right)^2 \right] \\
&\qquad\qquad+ 2 a^{-1} \left[ ( u^\mu + a k^\mu ) \Upsilon^{(1)} - U_{(1)}^\mu \right] \nabla_\mu \left( \gamma^{ab} \gamma^{(1)}_{ab} \right) + u^\mu \nabla_\mu \left( \gamma_{(1)}^{ab} \gamma^{(1)}_{ab} - 2 \gamma^{ab} \gamma^{(2)}_{ab} \right) \\
&\qquad\qquad- 4 a^{-1} \nabla_\mu U_{(2)}^\mu + 4 a^{-2} \Upsilon^{(1)} \nabla_\mu U_{(1)}^\mu + 2 a k^\mu \nabla_\mu \left( a^{-1} \Upsilon^{(1)} \right)^2 \bigg\} \eqend{,} \raisetag{1.8em}
\end{splitequation}
where all indices are raised and lowered with the background metric $g_{\mu\nu}$, and it is understood that $U_{(1)}^\mu$ and $\gamma_{\mu\nu}^{(1)}$ have only angular components:
\begin{equation}
u_\mu U_{(1)}^\mu = k_\mu U_{(1)}^\mu = 0 = u^\mu \gamma_{\mu\nu}^{(1)} = k^\mu \gamma_{\mu\nu}^{(1)} \eqend{.}
\end{equation}

It is interesting to compare this invariant Hubble rate with the one defined by the expansion of the constant-inflaton hypersurfaces. The normal vector field defining this foliation in the perturbed spacetime is
\begin{equation}
\label{eq:n}
\tilde{n}_\mu(x) \equiv \frac{\partial_\mu \tilde{\phi}(x)}{\sqrt{ - \tilde{g}^{\alpha\beta}(x) \partial_\alpha \tilde{\phi}(x) \partial_\beta \tilde{\phi}(x) }} \eqend{.}
\end{equation}
We recall that at the background level the geodesic observer co-moves with the inflaton, and this expression reduces to
\begin{equation}\label{eq:n_background}
n_\mu = u_\mu\eqend{.}
\end{equation}
The Hubble rate in this case is
\begin{equation}
\label{eq:H_phi_def}
\tilde{H}_\phi(x) \equiv \frac{1}{n-1} \tilde{\nabla}_\mu \tilde{n}^\mu(x) \eqend{,}
\end{equation}
and the corresponding invariant observable is defined by
\begin{equation}
\label{eq:H_phi_inv_def}
\mathcal{H}_\phi(\tilde{X}) \equiv \tilde{H}_\phi[x(\tilde{X})] \eqend{.}
\end{equation}
As in the previous case, it is convenient to first transform the normal vector field~\eqref{eq:n} to the field-dependent coordinates $\tilde{X}^{(\mu)}$. The result is
\begin{equation}
\label{eq:N}
\mathcal{N}_\mu(\tilde{X}) \equiv \frac{\partial_\mu \Phi(\tilde{X})}{\sqrt{ - \mathcal{G}^{\alpha\beta}(\tilde{X}) \partial_\alpha \Phi(\tilde{X}) \partial_\beta \Phi(\tilde{X}) }} \eqend{,}
\end{equation}
where we recall that $\Phi$ is the gauge-invariant inflaton field defined in Eq.~\eqref{eq:Phi}. Using the invariant normal vector $\mathcal{N}^\mu = \mathcal{G}^{\mu\nu} \mathcal{N}_\nu$, we can write Eq.~\eqref{eq:H_phi_inv_def} in the form
\begin{equation}\label{eq:H_phi_div}
\mathcal{H}_\phi(\tilde{X}) = \frac{1}{n-1} \bnabla_\mu \mathcal{N}^\mu(\tilde{X}) \eqend{.}
\end{equation}

To obtain the perturbative expansion of $\mathcal{H}_\phi$, we first expand the invariant normal vector $\mathcal{N}^\mu$ with respect to the inflaton perturbations. Using the expansion of the invariant inverse metric~\eqref{eq:inverse_inv_metric_hmunu_expansion}, the invariant inflaton~\eqref{eq:Phi} and the invariant four-velocity $\mathcal{U}^\mu$~\eqref{eq:U_inv_expansion}, together with the background value~\eqref{eq:n_background}, we obtain the perturbative expansion
\begin{splitequation}
\mathcal{N}^\mu &= \mathcal{U}^\mu - \frac{\kappa}{\dot\phi} \left( g^{\mu\nu} + u^\mu u^\nu \right) \nabla_\nu \left[ \Phi^{(1)} + \kappa \Phi^{(2)} \right] + \frac{\kappa^2}{\dot\phi} \Big( \mathcal{H}^{\mu\rho} + u_\nu u^\rho \mathcal{H}^{\mu\nu} + u^\mu u_\nu \mathcal{H}^{\nu\rho} \Big) \nabla_\rho \Phi^{(1)} \\
&\quad+ \frac{\kappa^2}{2 \dot\phi^2} \left[ u^\mu \left( g^{\alpha\beta} + 3 u^\alpha u^\beta \right) \nabla_\alpha \Phi^{(1)} \nabla_\beta \Phi^{(1)} + 2 \nabla^\mu \Phi^{(1)} u^\rho \nabla_\rho \Phi^{(1)} \right] + \bigo{\kappa^3} \eqend{.} \raisetag{2em}
\end{splitequation}
The invariant divergence of the invariant normal vector $\mathcal{N}^\mu$ can then be cast in the form
\begin{splitequation}
\bnabla_\mu \mathcal{N}^\mu &= \bnabla_\mu \mathcal{U}^\mu - \frac{\kappa}{a^2 \dot\phi} \nabla_\mu \left\{ \left( g^{\mu\nu} + u^\mu u^\nu \right) \nabla_\nu \left[ \Phi^{(1)} + \kappa \Phi^{(2)} \right] \right\} \\
&\quad+ \frac{\kappa^2}{\dot\phi} \nabla_\mu \left( g^{\alpha\beta} \mathcal{H}_{\alpha\beta} \right) \left( g^{\mu\nu} + u^\mu u^\nu \right) \nabla_\nu \Phi^{(1)} \\
&\quad+ \kappa^2 \nabla_\mu \left[ \frac{1}{\dot\phi} \left( \mathcal{H}^{\mu\rho} + \mathcal{H}^{\mu\nu} u_\nu u^\rho + u^\mu u_\nu \mathcal{H}^{\nu\rho} \right) \nabla_\rho \Phi^{(1)} \right] \\
&\quad+ \frac{\kappa^2}{2} \nabla_\mu \left[ \frac{1}{\dot\phi^2} \left( u^\mu g^{\alpha\beta} + 3 u^\mu u^\alpha u^\beta + 2 g^{\mu\alpha} u^\beta \right) \nabla_\alpha \Phi^{(1)} \nabla_\beta \Phi^{(1)} \right] + \bigo{\kappa^3} \eqend{.}
\end{splitequation}
We finally substitute Eqs.~\eqref{eq:H_covariant_components} and~\eqref{eq:H_contravariant_components} into this expression, which results in
\begin{splitequation}
\label{eq:H_phi_final}
\mathcal{H}_\phi &= \mathcal{H}_u - \frac{\kappa}{(n-1) a^2 \dot\phi} \left\{ \nabla_\mu \hat\nabla^\mu \left[ \Phi^{(1)} + \kappa \Phi^{(2)} \right] - a^2 \nabla_\mu \left( \frac{2}{a} \Upsilon^{(1)} + \gamma^{ab} \gamma^{(1)}_{ab} \right) \hat\nabla^\mu \Phi^{(1)} \right\} \\
&\quad+ \frac{\kappa^2}{n-1} \nabla_\mu \left\{ \frac{1}{a \dot\phi} \left[ 2 \Upsilon^{(1)} (u + ak)^{(\mu} (u + ak)^{\nu)} - 2 (u + ak)^{(\mu} U^{\nu)}_{(1)} + a \gamma^{\mu\nu}_{(1)} \right] \hat\nabla_\nu \Phi^{(1)} \right\} \\
&\quad+ \frac{\kappa^2}{2 (n-1)} \nabla_\mu \left\{ \frac{1}{\dot\phi^2} \left[ u^\mu \hat\nabla^\nu \Phi^{(1)} + 2 \hat\nabla^\mu \Phi^{(1)} u^\nu \right] \nabla_\nu \Phi^{(1)} \right\} + \bigo{\kappa^3} \eqend{,} \raisetag{2em}
\end{splitequation}
where $\mathcal{H}_u$ was given in Eq.~\eqref{eq:H_u_final}, and we have defined the spatial derivative
\begin{equation}
\hat\nabla^\mu \Phi^{(1)} \equiv \left( g^{\mu\nu} + u^\mu u^\nu \right) \nabla_\nu \Phi^{(1)} \eqend{.}
\end{equation}
We see that the invariant Hubble parameters $\mathcal{H}_u$ and $\mathcal{H}_\phi$ generally differ from each other, which was to be expected. However, the difference vanishes whenever the spatial derivative $\hat\nabla_\nu \Phi^{(1)}$ of the invariant inflation perturbation vanishes, \ie, whenever the observer sees spatially homogeneous constant-inflaton hypersurfaces.

\section{Discussion}                                                                                                                                %
\label{sec:discussions}                                                                                                                             %

We have used the approach of Refs.~\cite{brunetti_etal_jhep_2016,froeb_cqg_2018,froeb_lima_cqg_2018} to perturbatively construct a new class of gauge-invariant (relational) observables in cosmology adapted to measurements on the observer's past lightcone. The observables were defined by computing perturbative corrections to the geodesic lightcone (GLC) coordinates of Ref.~\cite{gasperini_et_al_jcap_2011}. The corrections were built out of the gauge-dependent degrees of freedom of the metric perturbation around a background FLRW spacetime in an arbitrary gauge. The full perturbed GLC coordinates we have obtained are functionals of the metric, and their residual freedom is completely fixed by requiring them to agree with the background GLC coordinates in the absence of perturbations. We have seen in Sec.~\ref{sec:perturbed_glc_coord} that the perturbative corrections to the GLC coordinates depend on the metric perturbation integrated along the observer's past lightcone. This makes the invariant observables we have constructed causal, albeit non-local in general. Nevertheless, since the metric perturbation is gauge-dependent, we have shown that one can choose a gauge where the perturbed and background GLC coordinates agree. In this non-linear GLC gauge, there are no corrections to the GLC coordinates, and the invariant observables that we have constructed become local.

The invariant observables were obtained by performing a field-dependent diffeomorphism from the background coordinates to the perturbed, field-dependent GLC coordinates. We have used this framework to find explicit expressions for invariant observables corresponding to the perturbed metric and to the local expansion rate of the spacetime, the local Hubble rate, in an arbitrary gauge for the metric perturbations and up to second order in perturbation theory. In the analysis of the invariant metric in Sec.~\ref{sec:gauge_invariant_glc_metric}, we have shown that the invariant metric has the form of a metric in the GLC gauge~\eqref{eq:glc_metric_perturbed}, and we have determined the explicit expression of its components in terms of the metric perturbation. We have also presented a new derivation of the linearised Einstein's equations in single-field inflation directly in terms of the metric perturbations in the GLC gauge, and written down the dynamical and constrain equations. We then compared the decomposition of the metric perturbation in the GLC gauge with the standard decomposition in the conformally-flat coordinates, obtaining the explicit relations between the gauge-invariant variables in these decompositions. Furthermore, we have shown how the famous Sasaki--Mukhanov variable can be written in terms of the GLC components of the invariant metric.

We have also worked out explicit expressions for two distinct definitions for the local Hubble rate, one measuring the expansion of the spatial section of the geodesic observers and the other measuring the expansion of the hypersurfaces of constant inflaton field. Up to second order, our results show that these two definitions for the Hubble rate differ by terms involving spatial derivatives (in the observer's frame) of the gauge-invariant inflaton perturbation. Thus, at the classical level, these observables coincide whenever the inflaton perturbation is homogeneous and isotropic. At the quantum level, however, it is possible that even with the fields in a homogeneous and isotropic state, and hence a spatially constant expectation value, fluctuations and the resulting correlations in the inflaton field give rise to a non-vanishing contribution for that difference, distinguishing between the two observables. We leave this issue to further work.

In their construction of invariant observables, the authors of Ref.~\cite{fanizza_etal_jcap_2021} used a diffeomorphism that is defined by the same differential equations as Eqs.~\eqref{eq:eom_1}, but with their integration conditions differing from ours --- see the discussion at the end of Sec.~\ref{sec:perturbed_glc_coord}. Moreover, their observables are defined with respect to a certain perturbed redshift proper time $\tilde{\tau}_z$. In order to define this time coordinate, we first write down the expression for the perturbed redshift $\tilde{z}$ as a function of the background time coordinate $\tau$ when the photon was emitted:
\begin{equation}
1 + \tilde{z}(\tau,w,\theta) = \frac{\tilde{Y}(\tau_\text{o},w,\theta)}{\tilde{Y}(\tau,w,\theta)} \eqend{,}
\end{equation}
where $\tau_\text{o}$ is the background time coordinate of the photon reception by the observer (which we assume to be fixed) and $\tilde{Y}$ was defined in Eq.~\eqref{eq:Y}. The coordinate $\tilde{\tau}_z$ at the source is then defined by the observed redshift as
\begin{equation}
\frac{a(\tau_\text{o})}{a[\tilde{\tau}_z(\tau,w,\theta)]} = \frac{\tilde{Y}(\tau_\text{o},w,\theta)}{\tilde{Y}(\tau,w,\theta)} \eqend{,}
\end{equation}
and is adapted to the observation of light signals propagating from a localised source to the observer. It is a straightforward exercise to convert the invariant observables from the coordinates that we have used in this work to use the observed perturbed redshift. On the other hand, the invariant observables in the coordinates that we have proposed here are more suited to probe the combined effect of perturbations all over the observer's past lightcone. This is the case if we are interested, \eg, in the backreaction of quantum fluctuations of the metric on the local Hubble rate. We hope to come back to this issue in the future.

\acknowledgments

M.F.\ thanks R.\ Durrer for discussions on the geodesic lightcone coordinates. We also thank G.\ Fanizza, G.\ Marozzi and M.\ Medeiros for correspondence, and the anonymous referee for questions, both of which have improved the manuscript. The work of M.F.\ was funded by the Deutsche Forschungsgemeinschaft (DFG, German Research Foundation) --- project Nos.\ 415803368 and 406116891 within the Research Training Group RTG 2522/1. The work of W.L.\ was supported by the grant No.\ RPG-2018-400, ``Euclidean and in-in formalisms in static spacetimes with Killing horizons'', from the Leverhulme Trust.

\clearpage

\bibliography{references_v6}

\providecommand{\href}[2]{#2}\begingroup\raggedright\begin{thebibliography}{10}

\bibitem{conley_etal_ajs_2011}
A.~Conley et~al., {\it {Supernova Constraints and Systematic Uncertainties from
  the First 3 Years of the Supernova Legacy Survey}},
  \href{http://dx.doi.org/10.1088/0067-0049/192/1/1}{{\em Astrophys.~J.~Suppl.}
  {\bf 192} (2011) 1}, [\href{http://arxiv.org/abs/1104.1443}{{\tt
  arXiv:1104.1443}}].

\bibitem{planck_2018a}
{\bf Planck} Collaboration, N.~Aghanim et~al., {\it {Planck 2018 results. VI.
  Cosmological parameters}},
  \href{http://dx.doi.org/10.1051/0004-6361/201833910}{{\em Astron.~Astrophys.}
  {\bf 641} (2020) A6}, [\href{http://arxiv.org/abs/1807.06209}{{\tt
  arXiv:1807.06209}}].

\bibitem{planck_2018b}
{\bf Planck} Collaboration, Y.~Akrami et~al., {\it {Planck 2018 results. IX.
  Constraints on primordial non-Gaussianity}},
  \href{http://dx.doi.org/10.1051/0004-6361/201935891}{{\em Astron.~Astrophys.}
  {\bf 641} (2020) A9}, [\href{http://arxiv.org/abs/1905.05697}{{\tt
  arXiv:1905.05697}}].

\bibitem{planck_2018c}
{\bf Planck} Collaboration, Y.~Akrami et~al., {\it {Planck 2018 results. X.
  Constraints on inflation}},
  \href{http://dx.doi.org/10.1051/0004-6361/201833887}{{\em Astron.~Astrophys.}
  {\bf 641} (2020) A10}, [\href{http://arxiv.org/abs/1807.06211}{{\tt
  arXiv:1807.06211}}].

\bibitem{lsstdark_energy_science_arxiv_2012}
{\bf LSST Dark Energy Science} Collaboration, A.~Abate et~al., {\it {Large
  Synoptic Survey Telescope: Dark Energy Science Collaboration}},
  \href{http://arxiv.org/abs/1211.0310}{{\tt arXiv:1211.0310}}.

\bibitem{euclid_th_working_grp_lrr_2013}
{\bf Euclid Theory Working Group} Collaboration, L.~Amendola et~al., {\it
  {Cosmology and fundamental physics with the Euclid satellite}},
  \href{http://dx.doi.org/10.12942/lrr-2013-6}{{\em Living Rev.~Rel.} {\bf 16}
  (2013) 6}, [\href{http://arxiv.org/abs/1206.1225}{{\tt arXiv:1206.1225}}].

\bibitem{desi_arxiv_2016}
{\bf DESI} Collaboration, A.~Aghamousa et~al., {\it {The DESI Experiment Part
  I: Science,Targeting, and Survey Design}},
  \href{http://arxiv.org/abs/1611.00036}{{\tt arXiv:1611.00036}}.

\bibitem{cmb-s4_arxiv_2016}
{\bf CMB-S4} Collaboration, K.~N. Abazajian et~al., {\it {CMB-S4 Science Book,
  First Edition}},  \href{http://arxiv.org/abs/1610.02743}{{\tt
  arXiv:1610.02743}}.

\bibitem{dodelson_cosmology_book}
S.~Dodelson, {\em {Modern Cosmology}}.
\newblock
  \href{http://www.worldcat.org/search?q=isbn:978-0-12-219141-1}{Academic
  Press, Amsterdam, The Netherlands, 2003}.

\bibitem{froeb_hack_higuchi_jcap_2017}
M.~B. Fr{\"o}b, T.-P. Hack, and A.~Higuchi, {\it {Compactly supported
  linearised observables in single-field inflation}},
  \href{http://dx.doi.org/10.1088/1475-7516/2017/07/043}{{\em JCAP} {\bf 7}
  (2017) 43}, [\href{http://arxiv.org/abs/1703.01158}{{\tt arXiv:1703.01158}}].

\bibitem{froeb_hack_khavkine_cqg_2018}
M.~B. Fr{\"o}b, T.-P. Hack, and I.~Khavkine, {\it {Approaches to linear local
  gauge-invariant observables in inflationary cosmologies}},
  \href{http://dx.doi.org/10.1088/1361-6382/aabcb7}{{\em Class.~Quantum~Grav.}
  {\bf 35} (2018) 115002}, [\href{http://arxiv.org/abs/1801.02632}{{\tt
  arXiv:1801.02632}}].

\bibitem{khavkine_cqg_2019}
I.~Khavkine, {\it {Compatibility complexes of overdetermined PDEs of finite
  type, with applications to the Killing equation}},
  \href{http://dx.doi.org/10.1088/1361-6382/ab329a}{{\em Class.~Quantum~Grav.}
  {\bf 36} (2019), no.~18 185012}, [\href{http://arxiv.org/abs/1805.03751}{{\tt
  arXiv:1805.03751}}].

\bibitem{torre_prd_1993}
C.~Torre, {\it {Gravitational observables and local symmetries}},
  \href{http://dx.doi.org/10.1103/PhysRevD.48.R2373}{{\em Phys.~Rev.~D} {\bf
  48} (1993) 2373}, [\href{http://arxiv.org/abs/gr-qc/9306030}{{\tt
  gr-qc/9306030}}].

\bibitem{giddings_marolf_hartle_prd_2006}
S.~B. Giddings, D.~Marolf, and J.~B. Hartle, {\it {Observables in effective
  gravity}},  \href{http://dx.doi.org/10.1103/PhysRevD.74.064018}{{\em
  Phys.~Rev.~D} {\bf 74} (2006) 064018},
  [\href{http://arxiv.org/abs/hep-th/0512200}{{\tt hep-th/0512200}}].

\bibitem{khavkine_cqg_2015}
I.~Khavkine, {\it {Local and gauge invariant observables in gravity}},
  \href{http://dx.doi.org/10.1088/0264-9381/32/18/185019}{{\em
  Class.~Quantum~Grav.} {\bf 32} (2015) 185019},
  [\href{http://arxiv.org/abs/1503.03754}{{\tt arXiv:1503.03754}}].

\bibitem{komar_pr_1958}
A.~Komar, {\it {Construction of a Complete Set of Independent Observables in
  the General Theory of Relativity}},
  \href{http://dx.doi.org/10.1103/PhysRev.111.1182}{{\em Phys.~Rev.} {\bf 111}
  (1958), no.~4 1182}.

\bibitem{bergmann_komar_prl_1960}
P.~G. Bergmann and A.~B. Komar, {\it {Poisson brackets between locally defined
  observables in general relativity}},
  \href{http://dx.doi.org/10.1103/PhysRevLett.4.432}{{\em Phys.~Rev.~Lett.}
  {\bf 4} (1960) 432}.

\bibitem{bergmann_rmp_1961}
P.~G. Bergmann, {\it {Observables in General Relativity}},
  \href{http://dx.doi.org/10.1103/RevModPhys.33.510}{{\em Rev.~Mod.~Phys.} {\bf
  33} (1961) 510}.

\bibitem{tambornino_sigma_2012}
J.~Tambornino, {\it {Relational Observables in Gravity: a Review}},
  \href{http://dx.doi.org/10.3842/SIGMA.2012.017}{{\em SIGMA} {\bf 8} (2012)
  017}, [\href{http://arxiv.org/abs/1109.0740}{{\tt arXiv:1109.0740}}].

\bibitem{brown_kuchar_prd_1995}
J.~D. Brown and K.~V. Kucha{\v r}, {\it {Dust as a standard of space and time
  in canonical quantum gravity}},
  \href{http://dx.doi.org/10.1103/PhysRevD.51.5600}{{\em Phys.~Rev.~D} {\bf 51}
  (1995) 5600}, [\href{http://arxiv.org/abs/gr-qc/9409001}{{\tt
  gr-qc/9409001}}].

\bibitem{bardeen_prd_1980}
J.~M. Bardeen, {\it {Gauge invariant cosmological perturbations}},
  \href{http://dx.doi.org/10.1103/PhysRevD.22.1882}{{\em Phys.~Rev.~D} {\bf 22}
  (1980) 1882}.

\bibitem{sasaki_ptp_1986}
M.~Sasaki, {\it {Large Scale Quantum Fluctuations in the Inflationary
  Universe}},  \href{http://dx.doi.org/10.1143/PTP.76.1036}{{\em
  Prog.~Theor.~Phys.} {\bf 76} (1986) 1036}.

\bibitem{mukhanov_zetf_1988}
V.~F. Mukhanov, {\it {Quantum theory of gauge-invariant cosmological
  perturbations}},  {\em Zh.~Eksp.~Teor.~Fiz.} {\bf 94} (1988) 1.
  [\href{http://jetp.ras.ru/cgi-bin/dn/e_067_07_1297.pdf}{Sov.~Phys.~JETP {\bf
  67} (1988) 1297}].

\bibitem{giesel_et_al_cqg_2010a}
K.~Giesel, S.~Hofmann, T.~Thiemann, and O.~Winkler, {\it {Manifestly
  Gauge-Invariant General Relativistic Perturbation Theory. I. Foundations}},
  \href{http://dx.doi.org/10.1088/0264-9381/27/5/055005}{{\em
  Class.~Quantum~Grav.} {\bf 27} (2010) 055005},
  [\href{http://arxiv.org/abs/0711.0115}{{\tt arXiv:0711.0115}}].

\bibitem{giesel_et_al_cqg_2010b}
K.~Giesel, S.~Hofmann, T.~Thiemann, and O.~Winkler, {\it {Manifestly
  Gauge-invariant general relativistic perturbation theory. II. FRW background
  and first order}},
  \href{http://dx.doi.org/10.1088/0264-9381/27/5/055006}{{\em
  Class.~Quantum~Grav.} {\bf 27} (2010) 055006},
  [\href{http://arxiv.org/abs/0711.0117}{{\tt arXiv:0711.0117}}].

\bibitem{giesel_et_al_arxiv_2020}
K.~Giesel, L.~Herold, B.-F. Li, and P.~Singh, {\it {Mukhanov-Sasaki equation in
  manifestly gauge-invariant linearized cosmological perturbation theory with
  dust reference fields}},
  \href{http://dx.doi.org/10.1103/PhysRevD.102.023524}{{\em Phys.~Rev.~D} {\bf
  102} (2020) 023524}, [\href{http://arxiv.org/abs/2003.13729}{{\tt
  arXiv:2003.13729}}].

\bibitem{brunetti_etal_jhep_2016}
R.~Brunetti, K.~Fredenhagen, T.-P. Hack, N.~Pinamonti, and K.~Rejzner, {\it
  {Cosmological perturbation theory and quantum gravity}},
  \href{http://dx.doi.org/10.1007/JHEP08(2016)032}{{\em JHEP} {\bf 08} (2016)
  032}, [\href{http://arxiv.org/abs/1605.02573}{{\tt arXiv:1605.02573}}].

\bibitem{froeb_cqg_2018}
M.~B. Fr{\"o}b, {\it {Gauge-invariant quantum gravitational corrections to
  correlation functions}},
  \href{http://dx.doi.org/10.1088/1361-6382/aaa74c}{{\em Class.~Quantum~Grav.}
  {\bf 35} (2018) 055006}, [\href{http://arxiv.org/abs/1710.00839}{{\tt
  arXiv:1710.00839}}].

\bibitem{froeb_lima_cqg_2018}
M.~B. Fr{\"o}b and W.~C.~C. Lima, {\it {Propagators for gauge-invariant
  observables in cosmology}},
  \href{http://dx.doi.org/10.1088/1361-6382/aab427}{{\em Class.~Quantum~Grav.}
  {\bf 35} (2018) 095010}, [\href{http://arxiv.org/abs/1711.08470}{{\tt
  arXiv:1711.08470}}].

\bibitem{froeb_cqg_2019}
M.~B. Fr{\"o}b, {\it {One-loop quantum gravitational backreaction on the local
  Hubble rate}},  \href{http://dx.doi.org/10.1088/1361-6382/ab10fb}{{\em
  Class.~Quantum~Grav.} {\bf 36} (2019) 095010},
  [\href{http://arxiv.org/abs/1806.11124}{{\tt arXiv:1806.11124}}].

\bibitem{lima_cqg_2021}
W.~C.~C. Lima, {\it {Graviton backreaction on the local cosmological expansion
  in slow-roll inflation}},
  \href{http://dx.doi.org/10.1088/1361-6382/abfaeb}{{\em Class.~Quantum~Grav.}
  {\bf 38} (2021) 135015}, [\href{http://arxiv.org/abs/2007.04995}{{\tt
  arXiv:2007.04995}}].

\bibitem{giesel_ijmpa_2008}
K.~Giesel, {\it {Introduction to Dirac observables}},
  \href{http://dx.doi.org/10.1142/S0217751X08040056}{{\em Int.~J.~Mod.~Phys.~A}
  {\bf 28} (2008) 1190}.

\bibitem{mukhanov_feldman_brandenberger_pr_1992}
V.~F. Mukhanov, H.~A. Feldman, and R.~H. Brandenberger, {\it {Theory of
  cosmological perturbations. Part 1. Classical perturbations. Part 2. Quantum
  theory of perturbations. Part 3. Extensions}},
  \href{http://dx.doi.org/10.1016/0370-1573(92)90044-Z}{{\em Phys.~Rept.} {\bf
  215} (1992) 203}.

\bibitem{temple_prsa_1938}
G.~Temple, {\it {New systems of normal co-ordinates for relativistic optics}},
  \href{http://dx.doi.org/10.1098/rspa.1938.0164}{{\em Proc.~R.~Soc.~A} {\bf
  168} (1938) 122}.

\bibitem{maartens_phdthesis_1980}
R.~Maartens, {\em {Idealised observations in relativistic cosmology}}.
\newblock PhD thesis, {University of Cape Town}, 1980.

\bibitem{ellis_etal_pr_1985}
G.~F.~R. Ellis, S.~D. Nel, R.~Maartens, W.~R. Stoeger, and A.~P. Whitman, {\it
  {Ideal observational cosmology}},
  \href{http://dx.doi.org/10.1016/0370-1573(85)90030-4}{{\em Phys.~Rep.} {\bf
  124} (1985) 315}.

\bibitem{maartens_matravers_cqg_1994}
R.~Maartens and D.~R. Matravers, {\it {Isotropic and semi-isotropic
  observations in cosmology}},
  \href{http://dx.doi.org/10.1088/0264-9381/11/11/011}{{\em
  Class.~Quantum~Grav.} {\bf 11} (1994) 2693}.

\bibitem{preston_poisson_prd_2006}
B.~Preston and E.~Poisson, {\it {Light-cone coordinates based at a geodesic
  world line}},  \href{http://dx.doi.org/10.1103/PhysRevD.74.064009}{{\em
  Phys.~Rev.~D} {\bf 74} (2006) 064009},
  [\href{http://arxiv.org/abs/gr-qc/0606093}{{\tt gr-qc/0606093}}].

\bibitem{gasperini_et_al_jcap_2011}
M.~Gasperini, G.~Marozzi, F.~Nugier, and G.~Veneziano, {\it {Light-cone
  averaging in cosmology: Formalism and applications}},
  \href{http://dx.doi.org/10.1088/1475-7516/2011/07/008}{{\em JCAP} {\bf 07}
  (2011) 008}, [\href{http://arxiv.org/abs/1104.1167}{{\tt arXiv:1104.1167}}].

\bibitem{fanizza_etal_jcap_2015}
G.~Fanizza, M.~Gasperini, G.~Marozzi, and G.~Veneziano, {\it {A new approach to
  the propagation of light-like signals in perturbed cosmological
  backgrounds}},  \href{http://dx.doi.org/10.1088/1475-7516/2015/08/020}{{\em
  JCAP} {\bf 08} (2015) 020}, [\href{http://arxiv.org/abs/1506.02003}{{\tt
  arXiv:1506.02003}}].

\bibitem{nugier_mg_2015}
F.~Nugier, {\it {The Geodesic Light-Cone Coordinates, an Adapted System for
  Light-Signal-Based Cosmology}},  in {\em {14th Marcel Grossmann Meeting on
  Recent Developments in Theoretical and Experimental General Relativity,
  Astrophysics, and Relativistic Field Theories}}, 2015.
\newblock \href{http://arxiv.org/abs/1508.07464}{{\tt arXiv:1508.07464}}.

\bibitem{fanizza_etal_jcap_2021}
G.~Fanizza, G.~Marozzi, M.~Medeiros, and G.~Schiaffino, {\it {The cosmological
  perturbation theory on the Geodesic Light-Cone background}},
  \href{http://dx.doi.org/10.1088/1475-7516/2021/02/014}{{\em JCAP} {\bf 02}
  (2021) 014}, [\href{http://arxiv.org/abs/2009.14134}{{\tt
  arXiv:2009.14134}}].

\bibitem{mitsou_etal_cqg_2021}
E.~Mitsou, G.~Fanizza, N.~Grimm, and J.~Yoo, {\it {Cutting out the cosmological
  middle man: General Relativity in the light-cone coordinates}},
  \href{http://dx.doi.org/10.1088/1361-6382/abd681}{{\em Class.~Quantum~Grav.}
  {\bf 38} (2021) 055011}, [\href{http://arxiv.org/abs/2009.14687}{{\tt
  arXiv:2009.14687}}].

\bibitem{nugier_jcap_2016}
F.~Nugier, {\it {From GLC to double-null coordinates and illustration with
  static black holes}},
  \href{http://dx.doi.org/10.1088/1475-7516/2016/09/019}{{\em JCAP} {\bf 09}
  (2016) 019}, [\href{http://arxiv.org/abs/1606.08296}{{\tt
  arXiv:1606.08296}}].

\bibitem{giddings_weinberg_prd_2020}
S.~Giddings and S.~Weinberg, {\it {Gauge-invariant observables in gravity and
  electromagnetism: black hole backgrounds and null dressings}},
  \href{http://dx.doi.org/10.1103/PhysRevD.102.026010}{{\em Phys.~Rev.~D} {\bf
  102} (2020), no.~2 026010}, [\href{http://arxiv.org/abs/1911.09115}{{\tt
  arXiv:1911.09115}}].

\bibitem{misner_thorne_wheeler_book}
C.~Misner, K.~Thorne, and J.~A. Wheeler, {\em {Gravitation}}.
\newblock
  \href{http://www.worldcat.org/search?q=isbn:9780716703440}{W.~H.~Freeman, San
  Francisco, USA, 1973}.

\bibitem{wald_gr_book}
R.~M. Wald, {\em {General Relativity}}.
\newblock \href{http://www.worldcat.org/search?q=isbn:0226870332}{The
  University of Chicago Press, Chicago, USA, 1984}.

\bibitem{fleury_nugier_fanizza_jcap_2016}
P.~Fleury, F.~Nugier, and G.~Fanizza, {\it {Geodesic-light-cone coordinates and
  the Bianchi I spacetime}},
  \href{http://dx.doi.org/10.1088/1475-7516/2016/06/008}{{\em JCAP} {\bf 06}
  (2016) 008}, [\href{http://arxiv.org/abs/1602.04461}{{\tt
  arXiv:1602.04461}}].

\bibitem{nugier_phdthesis_2013}
F.~Nugier, {\em {Lightcone Averaging and Precision Cosmology}}.
\newblock PhD thesis, {Universit{\'e} Pierre et Marie Curie - Paris VI}, 2013.

\bibitem{liddle_parsons_barrow_prd_1994}
A.~R. Liddle, P.~Parsons, and J.~D. Barrow, {\it {Formalizing the slow-roll
  approximation in inflation}},
  \href{http://dx.doi.org/10.1103/PhysRevD.50.7222}{{\em Phys.~Rev.~D} {\bf 50}
  (1994) 7222}, [\href{http://arxiv.org/abs/astro-ph/9408015}{{\tt
  astro-ph/9408015}}].

\bibitem{fanizza_etal_jacp_2013}
G.~Fanizza, M.~Gasperini, G.~Marozzi, and G.~Veneziano, {\it {An exact Jacobi
  map in the geodesic light-cone gauge}},
  \href{http://dx.doi.org/10.1088/1475-7516/2013/11/019}{{\em JCAP} {\bf 11}
  (2013) 019}, [\href{http://arxiv.org/abs/1308.4935}{{\tt arXiv:1308.4935}}].

\bibitem{fanizza_etal_jacp_2019}
G.~Fanizza, M.~Gasperini, G.~Marozzi, and G.~Veneziano, {\it {Observation
  angles, Fermi coordinates, and the Geodesic-Light-Cone gauge}},
  \href{http://dx.doi.org/10.1088/1475-7516/2019/01/004}{{\em JCAP} {\bf 01}
  (2019) 004}, [\href{http://arxiv.org/abs/1812.03671}{{\tt
  arXiv:1812.03671}}].

\bibitem{stewart_walker_prsla_1974}
J.~M. Stewart and M.~Walker, {\it {Perturbations of space-times in general
  relativity}},  \href{http://dx.doi.org/10.1098/rspa.1974.0172}{{\em
  Proc.~R.~Soc.~A} {\bf 341} (1974) 49}.

\bibitem{lifshitz_grg_2017}
E.~Lifshitz, {\it {Republication of: On the gravitational stability of the
  expanding universe}},
  \href{http://dx.doi.org/10.1007/s10714-016-2165-8}{{\em
  Gen.~Relativ.~Gravit.} {\bf 49} (2017) 18}. [J.~Phys.~(USSR) {\bf 10} (1946)
  116].

\bibitem{froeb_jcap_2014}
M.~B. Fr{\"o}b, {\it {The Weyl tensor correlator in cosmological spacetimes}},
  \href{http://dx.doi.org/10.1088/1475-7516/2014/12/010}{{\em JCAP} {\bf 12}
  (2014) 010}, [\href{http://arxiv.org/abs/1409.7964}{{\tt arXiv:1409.7964}}].

\bibitem{geshnizjani_brandenberger_prd_2002}
G.~Geshnizjani and R.~Brandenberger, {\it Back reaction and local cosmological
  expansion rate},  \href{http://dx.doi.org/10.1103/PhysRevD.66.123507}{{\em
  Phys.~Rev.~D} {\bf 66} (2002) 123507},
  [\href{http://arxiv.org/abs/gr-qc/0204074}{{\tt gr-qc/0204074}}].

\end{thebibliography}\endgroup

\end{document}